\documentclass[preprint,12pt,authoryear]{elsarticle}
\newpageafter{author}

\usepackage{amssymb}
\usepackage{amsmath}

\usepackage{xcolor}
\usepackage[%
 setpagesize=false,%
 bookmarks=true,%
 bookmarksnumbered=true,%
 colorlinks=true,%
 breaklinks=true,%
 pdftitle={},%
 pdfsubject={},%
 pdfauthor={},%
 pdfkeywords={}%
]{hyperref}

\begin{document}

\begin{frontmatter}



\title{Dynamic investment in disaster preparedness under recurrent and catastrophic risks} 


\author[ti]{Tomoki Ishikura \corref{cor1}} 
\ead{iskr@tmu.ac.jp}
\cortext[cor1]{Corresponding author}

\author[yoko]{Muneta Yokomatsu} 
\ead{yoko@iiasa.ac.at}

\affiliation[ti]{organization={Tokyo Metropolitan University},
            addressline={Minamiosawa 1-1}, 
            city={Hachioji},
            postcode={192-0397}, 
            state={Tokyo},
            country={Japan}}
\affiliation[yoko]{organization={International Institute for Applied Systems Analysis},
            addressline={Schlossplatz 1}, 
            city={Laxenburg},
            postcode={A-2361}, 
            country={Austria}}

\begin{abstract}
Long-term disaster risk reduction (DRR) requires investment decisions that account for uncertain disaster conditions and accumulated preparedness capacity. This study develops a novel stochastic dynamic control model for optimal DRR investment under recurrent and catastrophic uncertainty. The model combines diffusion and Poisson jump processes and treats preparedness capital as an endogenous stock that accumulates through investment, depreciates, and mitigates economic damage. We derive a state-dependent investment rule from the Hamilton–Jacobi–Bellman equation to minimize expected discounted investment expenditure and disaster damage over an infinite horizon.

Monte Carlo policy comparisons and numerical experiments examine how disaster risk structures interact with investment decisions to shape preparedness accumulation and economic losses. The results show that the economic value of DRR investment depends on adjustment to evolving conditions and disaster loss mechanisms. Lower expenditure can coexist with greater preparedness, while similar expected losses can conceal different upper-tail risks. Effective investment therefore requires attention to both accumulated preparedness and the structure of disaster risk.

These findings support long-term DRR planning that integrates investment timing, capital maintenance, and recovery conditions with residual loss assessment. Upper-tail indicators complement expected-loss evaluation without being directly optimized. The framework provides a basis for dynamically managing preparedness capital.

\end{abstract}




\end{frontmatter}


\color{black}

\section{Introduction}
\label{introduction}

Climate change, urbanization, and the concentration of economic activity in hazard-prone areas are changing both the magnitude and the structure of disaster risk. Societies must therefore decide not only how much to invest in disaster risk reduction (DRR), but also when to invest and how to build and maintain preparedness capacity over time. This problem is inherently dynamic because current DRR investment affects future preparedness, while preparedness benefits depend on uncertain future disaster conditions.

A further challenge is that disaster uncertainty takes different forms. Some hazards generate recurrent fluctuations in damage conditions, whereas others occur infrequently but cause catastrophic losses. Changes in exposure, disaster severity, event frequency, and recovery speed can further alter the economic consequences of these hazards. These mechanisms need not have equivalent implications for DRR policy. Even when two risk environments generate similar expected losses, they may differ substantially in the required accumulation of preparedness capital or in the upper tail of the loss distribution. Long-term DRR planning therefore requires a framework that relates investment decisions not only to expected disaster losses but also to the underlying structure of disaster risk.

Preparedness itself is also inherently dynamic. Infrastructure, institutional capacity, and other forms of preparedness are accumulated through past investment and depreciate over time. Consequently, current investment expenditure and the existing stock of preparedness capacity should be distinguished. A reduction in current expenditure does not necessarily imply declining preparedness if sufficient capacity has already been accumulated. Conversely, maintaining a constant expenditure level may be inefficient when disaster conditions and preparedness stocks evolve. Optimal DRR policy should therefore depend jointly on the current disaster state and the preparedness capacity inherited from previous investment.

Existing research has addressed important parts of this problem. Real-options approaches have examined the timing and flexibility of adaptation and flood-protection projects under uncertainty. Rare-disaster and catastrophe-risk models have shown the importance of low-probability, high-impact events for economic decision making. Dynamic disaster-economic models have, in turn, shown that DRR capital can generate long-term benefits that static project appraisal cannot capture. However, a gap remains between these strands of research. DRR investment is commonly represented as a discrete project decision, a predetermined expenditure rule, or an exogenous policy parameter, rather than as a continuous control that endogenously accumulates preparedness capacity. This limits our understanding of how the optimal investment path should adjust when recurrent uncertainty and catastrophic shocks coexist. Section 2 reviews these strands of literature and identifies this gap in detail.

This study addresses this gap by constructing a stochastic dynamic control model for long-term DRR policy. The methodological contribution lies in embedding recurrent disaster fluctuations and rare catastrophic shocks within a dynamic investment problem in which preparedness capital is an endogenous state variable and DRR investment is a continuous, state-dependent control. Recurrent fluctuations follow a diffusion process, while catastrophic events arrive through a Poisson jump process. Preparedness capital accumulates through investment, depreciates over time, reduces the flow of economic damage, and mitigates the magnitude of catastrophic damage. The resulting Hamilton–Jacobi–Bellman equation yields an optimal investment rule that links current investment to both the damage state and the accumulated preparedness stock. The model thereby formalizes the intertemporal trade-off between current preparedness expenditure and reductions in future economic losses.

The analysis contributes to long-term DRR planning in three important ways. Firstly, it offers a cohesive decision-making framework for understanding how to build preparedness in the face of both recurrent and catastrophic risks. Secondly, it assesses the economic value of state-dependent investment by contrasting the optimal policy with scenarios involving no additional investment and a fixed investment strategy. This comparison highlights the inherent benefits of preparedness investment as well as the added advantages of tailoring investment to changing disaster conditions. Lastly, the numerical experiments identify five key dimensions of disaster risk—recurrent uncertainty, growth in exposure, severity of catastrophic events, frequency of rare disaster events, and damage persistence—and investigate how each dimension influences optimal investment strategies, the accumulation of preparedness, and the long-term distribution of economic losses. Since the optimal policy aims to minimize expected discounted losses, the study also employs percentile and expected shortfall measures as post-evaluation diagnostics to analyze the remaining upper-tail exposure.

Through this framework, the study shifts the focus of DRR investment appraisal from isolated projects or predetermined expenditure levels toward the dynamic management of preparedness capacity under evolving disaster risk. The numerical results demonstrate that state-dependent investment can reduce expected and upper-tail losses relative to simpler investment rules. At the same time, different changes in the risk environment can produce markedly different combinations of investment expenditure, preparedness stocks, and residual losses. These findings indicate that neither current investment expenditure nor expected loss alone provides a sufficient basis for long-term DRR planning.

The remainder of this paper is organized as follows. Section 2 reviews the relevant literature. Section 3 develops the stochastic dynamic control model and derives the optimal investment condition. Section 4 presents the numerical solution, benchmark policy comparison, numerical experiments, and their policy implications. Section 5 concludes the paper.

\section{Literature review}
\label{literature review}


Climate change has substantially altered the characteristics of natural hazards, increasing the importance of DRR strategies that explicitly account for uncertain future conditions. 
Disaster impacts are not determined solely by physical hazard intensity but emerge from the interaction of hazards with socioeconomic systems encompassing exposure, vulnerability, infrastructure, and preparedness capacity. 
\cite{hallegatte2014} demonstrates that economic resilience — defined as the capacity of affected agents to absorb and recover from losses — is a key determinant of the macroeconomic consequences of disasters. 
\cite{noy2009} shows empirically that countries with higher literacy rates, stronger institutions, greater trade openness, and higher per capita incomes recover more rapidly from disaster shocks, and that output declines are significantly larger in developing nations following disasters of comparable physical magnitude (\cite{noy2009}, \cite{yokomatsu2024}). 
These findings underscore that post-disaster recovery capacity is at least as important as the reduction of hazard intensity in shaping long-term economic outcomes.  

These observations motivate a framework that treats disaster preparedness capacity as a dynamic stock accumulated through investment and depreciating over time, rather than a fixed exogenous variable. 
DRR investment has been shown to generate welfare effects even in the absence of actual disaster occurrence, because the mere reduction of destruction risk raises the shadow values of other productive assets and promotes broader capital accumulation (\cite{ishiwata2018}). 
This ex-ante risk reduction effect — distinct from the ex-post damage mitigation effect — is identifiable only within a stochastic dynamic optimization framework and is systematically underestimated by models that treat disaster as an unpredictable external shock. 
Disaster risk analysis therefore requires a framework that explicitly incorporates uncertainty and the dynamic evolution of preparedness capacity.


A substantial body of literature applies real options (RO) frameworks to investments in climate adaptation and flood protection. 
It emphasizes that such investments are irreversible, future conditions are uncertain, and delaying investment can itself have significant value.
\cite{gomez-cunya2020} develop the Evaluation of Investments in Flood Protection under Uncertainty (EIFU) framework, applying RO theory to determine the optimal timing of flood protection investment using historical river discharge data as a stochastic variable. 
They demonstrate that the traditional discounted cash flow (DCF) approach is inadequate for such decisions because it cannot accommodate the uncertain magnitude and timing of natural disasters, and that accounting for timing flexibility materially alters the investment decision. 
\cite{truong2018} extend this line of research by developing a real options framework for catastrophic risk reduction under climate change uncertainty, modeling the frequency of catastrophic events using a doubly stochastic Poisson process whose intensity evolves as a geometric Brownian motion. 
They find that the value of investment flexibility is large relative to the net present value (NPV) of considered projects, that losses from applying a simple NPV rule are especially severe when both uncertainty and project scale are large, and that it is generally optimal to invest in low sunk-cost projects first in order to preserve flexibility for future high-impact events. 
\cite{kim2018} apply RO analysis to flood control facility investment across 23 municipalities in the Nakdong River Basin, South Korea, demonstrating that managerial flexibility — particularly an expansion option — substantially improves economic feasibility under climate uncertainty, and emphasize that adaptation strategies should be grounded in future climate scenarios rather than historical data. 
\cite{segi2024} develop an RO model with a genetic algorithm solution method that dynamically optimizes budget allocation to flood control facilities under uncertain climate change scenarios, demonstrating that the flexible investment plan significantly reduces social costs relative to conventional inflexible planning. 
\cite{kind2018} apply RO analysis to flood risk management in the Netherlands, showing that while ROA offers valuable insights beyond standard cost-benefit analysis — particularly regarding high fixed costs and the benefits of adaptability — its practical implementation faces considerable challenges in recognizing uncertainty, quantifying scenarios, and integrating stakeholder preferences. 
\cite{corderi-novoa2021} develop a dynamic mixed integer linear programming model for flood risk reduction investment in a river basin, demonstrating that optimal investment is greater under higher rates of damage growth and under stronger aversion to extreme flood losses, while financing constraints affect only the rate of initial implementation.


Another line of research has explored decision-making in the face of rare but catastrophic risks.
\cite{barro2006} and \cite{barro2009} demonstrate that low-probability, high-impact events can substantially affect asset values, welfare, and optimal investment decisions, showing that rational agents would sacrifice a significant fraction of permanent consumption to eliminate the small probability of major economic collapse. 
\cite{gabaix2012variable} extends this framework by allowing time-varying disaster intensity, demonstrating that a variable rare disasters model can simultaneously resolve a broad set of macro-finance puzzles — including the equity premium, excess volatility, and return predictability. 
\cite{gourio2008} further shows that incorporating post-disaster recoveries into Barro-type models critically alters the implied equity premium, with the direction of the effect depending on the intertemporal elasticity of substitution. 
\cite{pindyck2013} develop a general equilibrium model in which catastrophic shocks arrive according to a Poisson process and destroy a random fraction of the capital stock, showing that conventional NPV analysis is insufficient because reducing catastrophic risk changes both cash flows and equilibrium asset prices. 
\cite{bretschger2019} apply a stochastic framework combining Wiener and Poisson processes to analyze optimal policy responses to environmental shocks, distinguishing among continuous fluctuations, discrete recurring jumps, and tipping points, and derive closed-form solutions for optimal growth paths under each regime. 
\cite{michel-kerjan2013} apply probabilistic catastrophe models to cost-benefit analysis of DRR investment in low- and middle-income countries, demonstrating that policymakers who base decisions solely on expected losses systematically undervalue low-probability extreme events, and that risk aversion is a material consideration in DRR investment appraisal. 
\cite{hong2023} examine subjective disaster probabilities formed through Bayesian learning by disaster-affected agents and derive optimal adaptation measures under climate-intensified disaster risks. 
These approaches clarify how environmental policy and adaptation can alter disaster risks, while leaving scope to examine continuous, state-dependent investment in a dedicated preparedness stock that depreciates over time and mitigates both ongoing damage and catastrophic shocks.

Dynamic optimization approaches have also been applied to DRR investment planning. 
\cite{altay2013} develop a dynamic model for disaster mitigation spending and show that the socially optimal level of mitigation shifts as governments accumulate knowledge and as economies become wealthier. 
\cite{yokomatsu2014} develop a dynamic stochastic macroeconomic model of DRR investment in developing countries, showing that DRR capital raises the shadow values of other productive assets by reducing the risk of destruction, and that this effect emerges even in periods in which no disaster actually occurs. 
\cite{ishiwata2018} formalize this insight by decomposing the total effect of DRR investment into the ex-ante risk reduction effect (ARRE) and the ex-post damage mitigation effect (PDME). Using a Pakistan case study, they demonstrate that ARRE accounts for approximately 40\% of the total DRR investment effect, and that models unable to capture ARRE systematically underestimate the value of DRR investment. 
They further explicitly identify, as a direction for future research, the need to define DRR investment as a control variable in the dynamic optimization problem — an approach that would describe a time-dependent optimal investment policy. 
\cite{yokomatsu2024} develop a dynamic model of climate-related disaster impacts with multidimensional household heterogeneity, demonstrating that DRR policy effects are nonmonotonic over time and depend critically on country-level macroeconomic context, yielding qualitatively different outcomes across economies characterized by different levels of human capital. 
The DYNAMMICs model (\cite{yokomatsu2022}) further demonstrates that infrastructure and risk-transfer investments generate co-benefits even in the absence of disaster occurrence, underscoring the importance of long-horizon dynamic evaluation.  

These studies establish the importance of dynamically accumulated DRR capital, while leaving scope to examine its accumulation and maintenance through continuous, state-dependent investment under coexisting recurrent fluctuations and catastrophic shocks.


The existing literature has made substantial progress along three dimensions. 
Real options approaches have demonstrated the value of investment flexibility and optimal timing for flood adaptation. 
Rare disaster economics and probabilistic catastrophe models have clarified why low-probability, high-impact events deserve special treatment in DRR investment appraisal. 
Dynamic macroeconomic DRR models have shown that preparedness capital generates ex-ante welfare effects that static, project-level, and shock-based models systematically fail to capture. Nevertheless, important gaps remain.  

First, stochastic environmental-economic models have already combined continuous fluctuations and catastrophic shocks \cite{bretschger2019}.
However, the implications of these coexisting risks for the optimal accumulation and maintenance of a dedicated stock of disaster preparedness capital remain insufficiently explored.
Second, existing DRR models often prescribe investment through fixed policy rates or exogenous rules, leaving scope to analyze investment as a continuous, state-dependent control that governs preparedness accumulation and future economic losses.
\cite{ishiwata2018} explicitly call for future work to define DRR investment as a control variable in the dynamic optimization problem, a direction this study directly pursues.
Third, existing studies rarely compare how qualitatively different risk environments — climate change, urbanization, and increasing disaster severity — generate different investment dynamics and long-term loss distributions, leaving policymakers without guidance on tailoring strategies to the specific character of local disaster risk.  

This study addresses these gaps through three contributions. 
We integrate frequent moderate hazard fluctuations and rare catastrophic shocks within a single stochastic framework, using a diffusion component for recurrent hazards and a Poisson jump component for catastrophic events. 
Our model represents preparedness as an endogenous capital stock that accumulates through investment, depreciates over time, and reduces both ongoing economic damage and the magnitude of catastrophic jumps in the damage state. Investment is determined as a continuous, state-dependent control within an infinite-horizon stochastic optimization problem. 
The numerical analysis further compares the optimal policy with no-investment and fixed-investment rules and examines how distinct risk mechanisms shape investment dynamics, preparedness accumulation, and long-term loss distributions. These comparisons assess the economic value of state-dependent investment adjustment and show why current expenditure, accumulated preparedness, and residual losses should be evaluated jointly.
Through these contributions, this study bridges the real options literature on climate adaptation investment, the economics of rare disasters, dynamic macroeconomic DRR modeling, and flood investment optimization, providing a unified analytical framework for long-term DRR investment planning under heterogeneous disaster risks.

\section{Model}
\label{Model}

\subsection{Assumptions and concepts}
\label{assumption}
This study treats the economic damage caused by disasters, together with DRR investments made during normal times, as economic losses arising from recurrent and catastrophic disaster risks
It constructs a mathematical programming model to manage policies that minimize these losses.
More specifically, we consider the expenditure associated with DRR investment and the economic damage incurred when a disaster occurs as the flow of economic losses over time. We then formulate the problem as a dynamic decision-making problem that minimizes the discounted present value of these losses over an infinite horizon, under uncertain disaster-occurrence risk.

The model defines two state variables: the level of disaster preparedness capital stock and the level of disaster damage at a given point in time.
The level of disaster preparedness capital stock does not refer solely to its quantitative magnitude, but rather to the extent of disaster preparedness and disaster mitigation functions. 
In other words, this concept encompasses not only the accumulated stock of physical capital, such as seawalls, levees, and earthquake-resistant facilities, but also intangible stocks that contribute to disaster preparedness and mitigation, including the degree of risk communication diffusion and social capital within local communities.
The level of disaster damage represents the degree of deviation from a fully sound state of the socioeconomic system in the current period. 
For example, it represents an aggregate index of conditions such as damage to housing and production facilities, dysfunction in infrastructure, including transportation networks, electricity, and water supply systems, and shortages in medical service capacity.
The damage flow in the economic system in each period is interpreted as all forms of disutility arising from the socioeconomic system being in an impaired state. 
It depends on the extent of the disaster's damage.

The economic system is exposed to disaster risk. 
This model considers two types of disaster risk.
The first consists of large-scale disasters that occur only rarely but cause severe damage. Examples include major earthquakes, volcanic eruptions, and extremely large typhoons.
The second consists of relatively high-frequency disasters that cause limited asset destruction and occur stochastically in each period. 
This type of risk is intended to represent the damage associated with recurrent events, such as heavy rainfall and droughts.

\subsection{Stochastic disaster process}
\label{stochastic process}

The stochastic disaster process consists of continuous fluctuations and discrete catastrophic shocks. 
The diffusion component represents frequent but moderate disaster variations, while the jump component captures rare catastrophic events. 
In addition, preparedness capital is modeled as an endogenous stock that accumulates through investment and reduces future disaster impacts. 
Table \ref{tab: risk component} summarizes the interpretation of each component in the proposed framework.

\begin{table}[htbp]
	\centering
	\caption{Interpretation of stochastic disaster risk components}
	\label{tab: risk component}

	\small

	\begin{tabular}{p{0.25\textwidth} p{0.32\textwidth} p{0.3\textwidth}}
	\hline
	\begin{tabular}{l}
		Model \\ component
	\end{tabular}	
	& \begin{tabular}{l}
		Disaster risk \\interpretation
	\end{tabular} & DRR implication \\
	\hline

	Diffusion component 
	($-\delta_N N_t dt + \beta N_t dz_t$)
	&
	Frequent and moderate disaster fluctuations: variations in rainfall
	conditions and recurrent flood risks
	&
	Routine disaster risk management and continuous preparedness needs
	\\[0.3cm]

	Jump component 
	($\phi {K_t}^{-\gamma}dq_t$)
	&
	Rare but extreme disaster events: catastrophic floods and
	low-probability high-impact hazards
	&
	Precautionary DRR investment against catastrophic risks
	\\[0.3cm]

	Preparedness capital 
	($K_t$)
	&
	Accumulated resilience capacity: preventive infrastructure,
	adaptation measures, and disaster preparedness activities
	&
	Long-term DRR investment that reduces future disaster impacts
	\\
	\hline
	\end{tabular}
\end{table}

Let $K_t$ and $x_t$ denote the disaster preparedness capital stock and DRR investment respectively.
In this study, preparedness capital $K$ represents an aggregated resilience capacity that includes physical infrastructure, institutional preparedness, and social capital.
DRR investment $x_t$ increases the disaster preparedness capital stock $K_t$, while the installation of capital stock entails adjustment costs (\cite{hayashi1982}).
The state equation for $K$ is written as:
\begin{equation}
	\label{definition: state equation of capital}
	dK_t = \left(x_t - \frac{\alpha}{2} {x_t}^2 -\delta_K K_t \right) dt ,
\end{equation}
where $\delta_K$ denotes a depreciation rate of $K$.
Following \cite{pindyck2013}, the model specifies adjustment costs as a quadratic function of the investment rate $x_t$, where $\alpha(>0)$ denotes the efficiency parameter with regard to the installation.
$h\left(K_t \right)$ represents the jump magnitude at which $N_t$ increases due to Poisson jumps, and the rate depends on the level of disaster preparedness capital stock $K_t$.

Let $N_t$ denote the level of disaster damage, whose law of motion is given by
\begin{equation}
	\label{definition: state equation of hazard}
	dN_t = -\delta_N N_t dt + \beta N_t dz_t + h(K_t) dq_t ,
\end{equation}
where $z_t$ is a standard Brownian motion, $\beta$ scales the diffusion term, and $q_t$ is a Poisson process.
Jumps arrive according to the Poisson process with a constant intensity $\mu$.
(\ref{definition: state equation of hazard}) implies that the level of disaster damage recovers at the rate $\delta_N$ each period, while it jumps upward by $h(K_t)$ when a rare disaster occurs.
$\delta_N$ captures the persistence and recovery of disaster risk under changing climatic conditions.

The function $h$ represents a general function capturing the idea that the damage increment per event of $N$ caused by rare disasters is mitigated by the accumulation of $K$.
In this study, we specify it as
\begin{equation}
	\label{mitigation of Jump}
	h(K) = \frac{\phi }{K^{\gamma}} \quad (\gamma > 0).
\end{equation}
$\gamma$ denotes a parameter that measures the contribution to disaster damage mitigation, while $\phi$ denotes the scale of the economy exposed to the hazard.
The parameter $\gamma$ represents the effectiveness of DRR investment in reducing disaster losses.

This model defines the current-period flow of economic losses as the sum of economic damage that depends on the disaster damage level and the investment expenditures.
We assume a current economic loss function is represented as 
\begin{equation}
	\label{definition: current loss function}
	U\left(x_t, K_t, N_t \right) =x_t + {N_t}^{\eta_N} {K_t}^{-\eta_K} \quad (\eta_N >0, \eta_K > 0).
\end{equation}

The current-period economic loss function $U$ defined in (\ref{definition: current loss function}) consists of two additive components, each capturing a distinct source of economic burden.
The first term, $x_t$, represents the flow of DRR investment in period $t$. 
This reflects the direct cost of building disaster preparedness capital $K_t$, including expenditures on physical infrastructure, institutional capacity, and intangible preparedness stocks. 
Even in the absence of a disaster event, this cost is incurred as long as the investment is undertaken.

The second term, $N_t^{\eta_N} K_t^{-\eta_K}$, represents the flow of economic damage attributable to the current state of disaster damage $N_t$, mitigated by the accumulated level of disaster preparedness capital $K_t$. Several structural features of this term deserve attention.

The damage flow is increasing and convex in $N_t$ when $\eta_N>1$, implying that, for a given preparedness stock, economic damage increases more than proportionally with the damage state.
This captures the notion that severe damage generates cascading disutility---for instance, simultaneous dysfunction
of transportation, utilities, and medical services compounds economic losses beyond their simple sum.

The damage flow is decreasing in $K_t$, with elasticity $\eta_K > 0$. This formalizes the core mechanism of disaster preparedness: a larger stock of preparedness capital---whether seawalls, earthquake-resistant facilities, or community-level risk communication---attenuates the economic consequences of a given damage state $N_t$. Importantly, $K_t$ enters multiplicatively with $N_t$, meaning that preparedness capital is most valuable precisely when disaster damage is high.

The two elasticity parameters $\eta_N$ and $\eta_K$ govern the curvature and sensitivity of the damage flow with respect to each state variable. A higher $\eta_N$ implies that the damage flow responds more sensitively to worsening disaster conditions, while $\eta_K$ determines how effectively preparedness capital suppresses damage costs. The asymmetry between these parameters allows the model to capture a wide range of real-world damage-mitigation relationships.

In sum, (\ref{definition: current loss function}) embeds a fundamental trade-off: reducing the second term requires accumulating $K_t$, which in turn necessitates incurring the first term. 
The specification of the loss function in (\ref{definition: current loss function}) warrants further justification. The power-law form $N_t^{\eta_N} K_t^{-\eta_K}$ belongs to the class of constant-elasticity damage functions widely
adopted in the disaster economics literature, because the elasticity parameters $\eta_N$ and $\eta_K$ are directly interpretable and amenable to empirical estimation (\cite{kousky2014informing},\cite{mechler2016disaster}).

The additive structure of (\ref{definition: current loss function}) reflects the institutional reality that preparedness expenditure and disaster damage are borne through distinct budgetary channels: the investment term $x_t$ represents deliberate ex-ante expenditures on resilience capacity, whereas $N_t^{\eta_N} K_t^{-\eta_K}$ represents ex-post economic losses incurred as a consequence of the realized disaster state. 
Treating these two components as additively separable is therefore consistent with the way in which DRR budgets and disaster loss accounts are recorded and evaluated in practice \cite{mechler2016disaster}, and allows the fundamental trade-off between investment cost and damage reduction to be analyzed in a transparent manner.

\subsection{Stochastic control and HJB equation}
\label{control and HJB}

We consider a social planner's problem of choosing DRR investment in each period to minimize the present value of economic losses over an infinite horizon, subject to a nonnegativity constraint on investment:

\begin{align}
	\label{Specific form of the model}
	& \min_{\boldsymbol{x} } \mathbb{E} \left\lbrace \int_{0}^{\infty} 
	\left(x_t + {N_t}^{\eta_N} {K_t}^{-\eta_K} \right) e^{-\rho t} dt \right\rbrace\\
	& s.t. \notag \\
	& x_t \ge 0, 
\end{align}
where $\rho$ denotes a constant rate of time preference and $\mathbb{E}$ denotes the expectation operator.

Let $V(K,N)$ denote the value function of this problem.
The Hamilton-Jacobi-Bellman (HJB) equation for the social planner is
\begin{equation}
	\label{HJB simple}
	\begin{split}
		\rho V\left(K,N \right)=& \min_{x} 
		\biggl[
		\left(x + N^{\eta_N} K^{-\eta_K} \right) \\
		& + \left( x - \frac{\alpha}{2} {x^2} -\delta_K K \right)
		\frac{\partial V}{\partial K}
			- \delta_N N 
			\frac{\partial V}{\partial N} \\
		&+ \mu \left\{ 
			V\left(
				K,N + \frac{\phi }{K^{\gamma}} 
			\right)
			-V\left(K,N\right) 
		\right\} \\
		&+ \left(
			\frac{\beta^2 N^2}{2}\frac{\partial^2 V}{\partial N^2}
		\right) \biggr].
	\end{split}
\end{equation}
For the proof, see \ref{app 1}.
We derive necessary conditions for a solution to the optimization problem on the right-hand side of (\ref{HJB simple}), taking into account the nonnegativity constraint on $x_t$. 
Specifically, a social planner chooses DRR investment in each period so as to minimize the present value of economic losses over an infinite horizon, subject to the nonnegativity constraint on investment.
Let $\mathcal{L}$ denote the Lagrangian of this problem, and $\nu(\ge 0)$ denote the Lagrange multiplier associated with the nonnegativity constraint on $x_t$, 
\begin{equation}
	\label{Lagrange}
	\begin{split}
		\mathcal{L}\left(x,\nu \right)=&\left(
			x + N^{\eta_N} K^{-\eta_K} 
		\right) \\
		&+ \left\lbrace
		\left( x - \frac{\alpha}{2} {x^2} -\delta_K K \right)
		\frac{\partial V}{\partial K}
		+ \left(
				-\delta_N N 
			\right)
			\frac{\partial V}{\partial N}		\right\rbrace \\
		&+ \mu \left\{ 
			V\left(
				K,N + \frac{\phi }{K^{\gamma}}
			\right)
			-V\left(K,N\right) 
		\right\} \\ 
		&+ \left(
			\frac{\beta^2 N^2}{2}\frac{\partial^2 V}{\partial N^2}
		\right)  \\
		&- \nu x .
	\end{split}
\end{equation}
Karush-Kuhn-Tucker (KKT) conditions for this problem are given by
\begin{align}
	\label{KKT linear}
	\frac{\partial \mathcal{L}}{\partial x}
	&= 1 + \left(1- \alpha x \right)\frac{\partial V}{\partial K} -\nu = 0\\
	\label{complementary}
	\nu x &=0 .
\end{align}
Thus, the above conditions yield the optimal DRR investment expenditure $x^{\ast}$ as
\begin{equation}
	\label{derivation optim x}
	x^{\ast} = \max \left\lbrace 
		\frac{1}{\alpha} \left(1 + \frac{1}{\frac{\partial V}{\partial K}}\right),0
	\right\rbrace .
\end{equation}
Focusing on the case in which $x^{\ast}>0$ in equation (\ref{derivation optim x}), we have
\begin{equation}
	\label{consider: optim x}
	x^{\ast} = \frac{1}{\alpha} 
		\left(1 + \left(\frac{\partial V}{\partial K}\right)^{-1} \right).
\end{equation}
Since $-\frac{\partial V}{\partial K}$ represents the marginal effect of an increase in the disaster preparedness capital stock on the reduction in total intertemporal economic losses, (\ref{consider: optim x}) shows that a larger marginal effect leads to a higher level of optimal DRR investment.
In other words, the mechanism determines the optimal level of DRR investment expenditure by balancing the present value of future disaster damage reduction achieved through DRR investment against current investment expenditure.

\section{Numerical analysis}
\label{section simulation}

\subsection{Benchmark setup}
\label{benchmark}

We conduct numerical simulations using the model constructed above. Since the HJB equation is a nonlinear partial differential equation with a nonlocal jump term, one generally cannot analytically derive the trajectory of $x^{\ast}$, except under special functional forms. 
Therefore, this study must rely on numerical analysis to examine the model's properties.

Accordingly, this study employs an iteration method that extends the collocation method proposed by \cite{judd1998} and \cite{miranda2002}. 
Specifically, this method approximates the value function by a polynomial function whose shape can flexibly vary with its parameters.
It then estimates the approximating function so that the HJB equation holds at discrete grid points in the continuous space spanned by the state variables.
For the detailed computational procedure, see \ref{app 2}.

We employ two-dimensional Chebyshev polynomials to approximate $V(K,N)$. 
Chebyshev polynomials constitute a class of orthogonal polynomials and are widely used for function approximation. 
Interpolation using the approximating function must be evaluated over a specified state space. 
In the present numerical simulation, the state space is set to $10 \le K \le 100$ and $0 \le N \le 30$. 
The simulation assigns 100 Chebyshev interpolation nodes to each of the ranges of $K$ and $N$, and estimates the polynomial coefficients to minimize the sum of squared HJB residuals at the resulting $100^2$ collocation points.
A Chebyshev node defines the $k$-th grid point for a continuous variable $y$ on the closed interval $[a,b]$, divided into $n$ nodes, as 
\begin{equation} 
	\label{chebyshev node} 
	y_k=\frac{1}{2}(a+b)+\frac{1}{2}(b-a) \cos \left(\frac{2k-1}{2n} \pi\right) 
\end{equation} 
and is known as a node placement that minimizes interpolation errors (\cite{judd1998}).

To characterize a hypothetical region exposed to disaster risk as a benchmark case, this study sets the model parameters as shown in Table \ref{table:parameter set}.
The parameter values reported in Table \ref{table:parameter set} are selected to represent a hypothetical region exposed to disaster risk, and are intended to be broadly consistent with the orders of magnitude found in the empirical and theoretical literature on disaster economics.
In this simulation, one period corresponds to one year.

The adjustment cost efficiency parameter $\alpha = 0.2$ implies that the marginal cost of investment rises moderately with the level of expenditure, reflecting the practical constraints on the rapid expansion of preparedness capacity such as procurement bottlenecks and institutional absorption limits. 
The depreciation rate of preparedness capital $\delta_K = 0.01$ corresponds to an e-folding time of 100 years in the absence of investment, which is broadly consistent with the long-run durability of major disaster preparedness infrastructure such as seawalls and levees.
an e-folding time of 100 years in the absence of investment

The damage recovery rate $\delta_N=0.2$ represents a continuous-time proportional recovery rate of 0.2 per year.
In the absence of diffusion and jump shocks, the damage state follows $N_t=N_0e^{-0.2t}$, implying an e-folding time of five years.
This is broadly consistent with empirical observations of long-term recovery processes reported in the disaster economics literature (\cite{noy2009}, \cite{hallegatte2014}).

The diffusion volatility and Poisson jump intensity are informed by Japanese precipitation and flood-damage data, using the common observation period of 2002--2022. 
One model period corresponds to one year.
These data provide empirical reference values for the uncertainty parameters of a hypothetical economy, rather than a structural estimation of the damage-state process.

For the diffusion component, we use the annual number of occurrences of hourly precipitation of at least 50~mm, recorded by the Japan Meteorological Agency's AMeDAS network and standardized to 1,300 observation stations (\cite{JMA2024}).
Let $R_t$ denote this annual precipitation-frequency indicator.
The sample standard deviation of its annual logarithmic changes, $\ln(R_t/R_{t-1})$, is approximately 0.282 over the observation period.

We therefore set $\beta=0.28$, assuming that the indicator’s relative annual variability provides a reference magnitude for the diffusion-driven fluctuations in $N$.
This assumption matches the scale of annual variability without requiring the precipitation indicator itself to follow the model's stochastic process.

For the jump component, we use the combined annual nominal flood damage to general assets and public civil engineering facilities reported in the Ministry of Land, Infrastructure, Transport and Tourism's Flood Damage Statistics (\cite{MLIT2022FloodDamage}).
We identify high-loss years using the upper outlier threshold $Q_{0.75}+1.5(Q_{0.75}-Q_{0.25})$, where $Q_p$ denotes the sample $p$-quantile of the combined damage series.
The resulting threshold is approximately JPY~957.2~billion, exceeded in 2004, 2018, and 2019.
Treating these three high-loss years among 21 observed years as proxies for years with at least one catastrophic event gives
an annual occurrence probability of $3/21$.
The corresponding Poisson intensity is $-\ln(1-3/21)\approx0.154$ per year, which we round to $\mu=0.15$.
This intensity implies a mean interarrival time of approximately 6.7~years and an annual probability of at least one jump of
approximately 13.9\%.

This calibration uses national aggregate indicators to characterize uncertainty in the hypothetical economy.
Annual damage totals do not identify individual disaster arrivals, and the classification of high-loss years remains affected by prices, exposure, and vulnerability.
Accordingly, the resulting parameters are interpreted as proxy-based benchmark values rather than direct estimates of regional hazard frequencies or the latent damage-state dynamics.

\color{black}
The rate of time preference $\rho = 0.04$ corresponds to the value of the social discount rate conventionally used in cost–benefit analyses of public investment projects in Japan.
The economic scale parameter $\phi = 10$ determines the baseline magnitude of the damage jump upon the occurrence of a rare disaster and is normalized to a representative order of magnitude for the hypothetical region. 
The elasticity parameters $\eta_N = 1$ and $\eta_K = 0.25$ govern the sensitivity of the damage flow to the damage level and to preparedness capital, respectively. 
The benchmark value $\eta_N = 1$ implies a linear relationship between the damage state and the flow of economic losses, while $\eta_K = 0.25$ implies that a one-percent increase in preparedness capital reduces the damage flow by 0.25 percent, reflecting a moderate but meaningful role for DRR investment in attenuating ongoing losses.

\begin{table}[t]
	\centering
	\caption{Parameter set for the benchmark case}
	\label{table:parameter set}
	\begin{tabular}{cc}
		\hline
		parameter & value \\ \hline
		$\alpha$ & 0.2 \\
		$\delta_K$ & 0.01 \\
		$\delta_N$ & 0.2 \\
		$\gamma$ & 0.2 \\
		$\eta_N$ & 1 \\
		$\eta_K$ & 0.25 \\
		$\beta$ & 0.28 \\
		$\rho$ & 0.04 \\
		$\mu$ & 0.15 \\
		$\phi$ & 10 \\
		\hline
	\end{tabular}
\end{table}

\begin{figure}
	\begin{minipage}{0.5\linewidth}
		\centering
		\includegraphics[bb=0 0 283 226, width=0.975\textwidth]{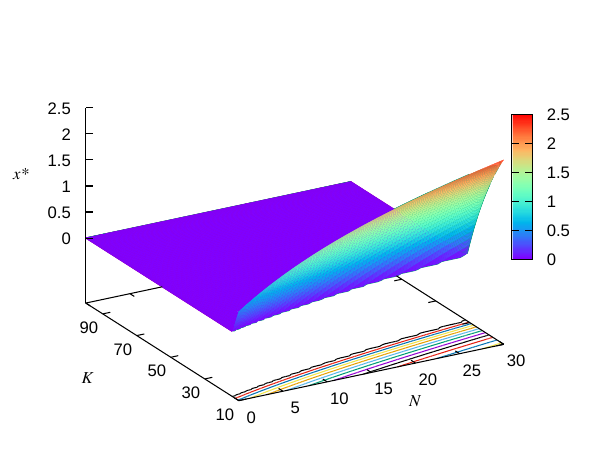}
		\caption{Optimal investment}
		\label{fig:case0 policy function}
	\end{minipage}
	\begin{minipage}{0.5\linewidth}
		\centering
		\includegraphics[bb=0 0 283 226, width=0.975\textwidth]{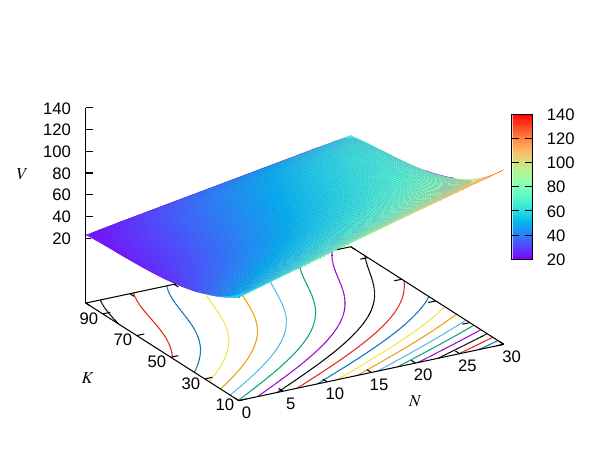}
		\caption{Value function}
		\label{fig:case value function}
	\end{minipage}	
\end{figure}

We summarize the properties of the derived optimal investment in disaster preparedness in the benchmark case. 
Figure \ref{fig:case0 policy function} shows that $x^{\ast}$ requires greater DRR investment when the disaster damage level $N$ is higher, and the stock of disaster preparedness capital $K$ is lower, which is consistent with intuition.
In addition, $x^{\ast}$ becomes zero in the region where $K$ is sufficiently large and $N$ is sufficiently small. According to Eq. (\ref{derivation optim x}), this corresponds to a state in which the marginal effect of increasing the stock of disaster preparedness capital is extremely small. 
In other words, when the economic system remains in sound condition and has accumulated sufficient preparedness capital, it does not need to undertake DRR investments.

The value function, that is, the present value of total economic losses over time, is small when $K$ is large and $N$ is small, as shown in Figure \ref{fig:case value function}.
This result is intuitive: a large stock of disaster preparedness capital $K$ suppresses the damage flow $N_t^{\eta_N} K_t^{-\eta_K}$ in each period, while a low damage level $N$ directly reduces the magnitude of that flow.
Together, these two conditions minimize the stream of economic losses that the social planner must bear over the infinite horizon.

Conversely, $V(K, N)$ becomes large when $K$ is small, and $N$ is large.
In this region of the state space, the economic system simultaneously suffers from high ongoing disaster damage and lacks sufficient preparedness capital to mitigate it. The social planner must therefore incur both a serious damage flow and the investment expenditure required to accumulate $K$, resulting in a large present value of total losses.

It is also worth noting the shape of the contours of $V(K, N)$. 
The rate at which $V$ decreases as $K$ increases is most pronounced when $N$ is large, reflecting the multiplicative structure of the damage term $N_t^{\eta_N} K_t^{-\eta_K}$ in (\ref{definition: current loss function}): disaster preparedness capital is most valuable ---in the sense of generating the greatest reduction in discounted future losses--- precisely when disaster damage is severe. 
This observation provides the intertemporal foundation for the policy function shown in Figure \ref{fig:case0 policy function}, where the optimal investment $x^*$ is highest in the region of high $N$ and low $K$.

\subsection{Benchmark policy comparison}
This subsection evaluates the significance of the optimal policy derived in Section~\ref{benchmark} by comparing it with simple investment rules. Hereafter, this policy is referred to as “Optimal.” All simulations in this subsection start from the initial conditions $K_0 = 15$ and $N_0 = 2$ and cover a 50-year horizon.

The simulation horizon reflects the 50-year post-completion period used in Japanese flood control project evaluations. 
These evaluations cover the construction period and the subsequent 50 years when calculating benefits and costs, including maintenance costs, and include the residual value of flood control assets at the end of the evaluation period in total benefits.

We compare Optimal with two alternative investment rules. Under the ``No Investment'' rule, no additional DRR investment is undertaken. 
Comparison with this rule assesses the effectiveness of DRR investment relative to relying solely on the initial stock of preparedness capital. 
Under the ``Fixed Investment'' rule, the investment rate remains constant over time at approximately 0.2, based on the average investment rate across all time intervals and all 500 simulation runs under Optimal. 
Comparison with this rule assesses the value of adjusting investment to the current state of the system.

We run 500 Monte Carlo simulations for each policy and compute indicators of discounted economic losses and their distributions.
In each simulation run, all three policies are assessed using the same realizations of the Poisson and Wiener processes. 
The continuous-time model is simulated with a time step of $\Delta t = 0.1$ years, dividing the 50-year horizon into 500 time intervals. 
To account for losses that extend beyond the simulation horizon, we incorporate the discounted continuation value, \( e^{-\rho T} V(K_T, N_T) \), into the total discounted losses accrued during the simulation, with \( T=50 \). 
This continuation value reflects the expected discounted losses from the terminal state over the remaining infinite horizon, assuming the optimal policy is adhered to after year 50 across all three scenarios.

The Fixed Investment rule maintains a constant investment rate equal to the mean investment rate under the Optimal policy. 
This mean is calculated across all 500 time intervals and all 500 Monte Carlo simulation runs:
\begin{equation}
    \label{eq:fixed_investment_rate}
    \bar{x}
    =
    \frac{1}{MJ}
    \sum_{m=1}^{M}
    \sum_{j=0}^{J-1}
    x^*(K_{t_j}^{(m)},N_{t_j}^{(m)}),
    \qquad M=J=500.
\end{equation}
The resulting annual investment rate is $\bar{x}=0.2021991$, approximately 0.2.
The Fixed Investment simulation uses the unrounded mean throughout. 
With $\Delta t=0.1$ years, the expenditure incurred during each interval is
$\bar{x}\Delta t\approx0.02022$.
This setting matches the average undiscounted investment expenditure over the 50-year horizon between Optimal and Fixed Investment, although their discounted expenditures can differ because of differences in investment timing.

Within each Monte Carlo run, all three policies use the same realizations of the Wiener and Poisson processes, while different realizations are used across runs. 
This allows comparison under identical disaster arrival times and stochastic fluctuations, making differences attributable to the investment rules easier to identify. 
However, the magnitude of each disaster-induced jump depends on preparedness capital through $h(K_t)$. 
Consequently, the same disaster arrival does not generate identical damage states under the three policies.

Let $t_j=j\Delta t$, where $\Delta t=0.1$ years and $j=0,\ldots,500$. 
In simulation run $m$, the Wiener increment over the interval $[t_j,t_{j+1}]$ is generated as
\begin{equation}
    \label{eq:benchmark_wiener_increment}
    \Delta W_j^{(m)}
    =
    W_{t_{j+1}}^{(m)}-W_{t_j}^{(m)}
    =
    \sqrt{\Delta t}\,\varepsilon_j^{(m)},
    \qquad
    \varepsilon_j^{(m)}\sim\mathcal{N}(0,1),
\end{equation}
where the standard normal draws are independent across time intervals and simulation runs. 
Thus, $\Delta W_j^{(m)}$ has mean zero and variance $\Delta t$. 
We use the same sequence of Wiener increments and the same Poisson process realization for all three policies within each run.

Let $\pi$ denote an investment policy and let $(x_t^\pi, K_t^\pi, N_t^\pi)$ denote the corresponding investment and state trajectories. 
The discounted economic losses accumulated over the 50-year simulation horizon are defined as 
\begin{equation}
    \label{eq:benchmark_loss_50}
    L_{50}^{\pi}
    =
    \int_0^{T}
    e^{-\rho t}
    \left[
        x_t^\pi
        +
        (N_t^\pi)^{\eta_N}(K_t^\pi)^{-\eta_K}
    \right]dt,
    \qquad T=50.
\end{equation}
The first term represents DRR investment expenditure, and the second represents economic damage. 
Both are expressed as flows per unit of time and discounted at the continuous-time rate $\rho$.
In the numerical simulation, this integral is approximated by
\begin{equation}
    \label{eq:benchmark_loss_discrete}
    L_{50}^{\pi}
    \approx
    \sum_{j=0}^{499}
    e^{-\rho t_j}
    \left[
        x_{t_j}^\pi
        +
        (N_{t_j}^\pi)^{\eta_N}(K_{t_j}^\pi)^{-\eta_K}
    \right]\Delta t.
\end{equation}
In particular, $x_{t_j}^\pi\Delta t$ is the investment expenditure incurred during one time interval.

To account for losses beyond the simulation horizon, we assume that the optimal policy is followed after year 50 under all three cases.
Let $V^*(K, N)$ denote the value function associated with the optimal policy. 
The terminal continuation value, expressed in present-value terms at time zero, is
\begin{equation}
    \label{eq:benchmark_terminal_value}
    C_{50}^{\pi}
    =
    e^{-\rho T}V^*(K_T^\pi,N_T^\pi),
    \qquad T=50.
\end{equation}
Here, $V^*(K_T^\pi, N_T^\pi)$ represents the expected discounted investment expenditures and economic damage over the remaining infinite horizon, conditional on the terminal state and evaluated at time $T$. 
Multiplication by $e^{-\rho T}$ discounts this continuation value to time zero.

The total discounted loss measure used in the policy comparison is therefore
\begin{equation}
    \label{eq:benchmark_total_loss}
    Q^\pi
    =
    L_{50}^\pi+C_{50}^\pi.
\end{equation}
This measure combines the losses realized during the first 50 years with the expected continuation losses under the optimal policy thereafter.

\begin{figure}
	\centering
	\includegraphics[bb=0 0 384 271, width=0.75\textwidth]{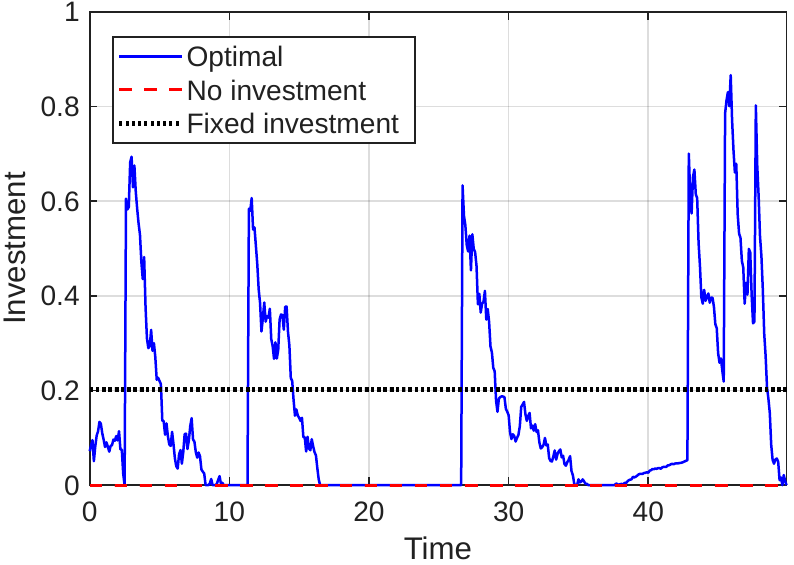}
	\caption{Sample paths of DRR investment under the three policies}
	\label{fig:sample x}
\end{figure}

\begin{figure}
	\centering
	\includegraphics[bb=0 0 384 271, width=0.75\textwidth]{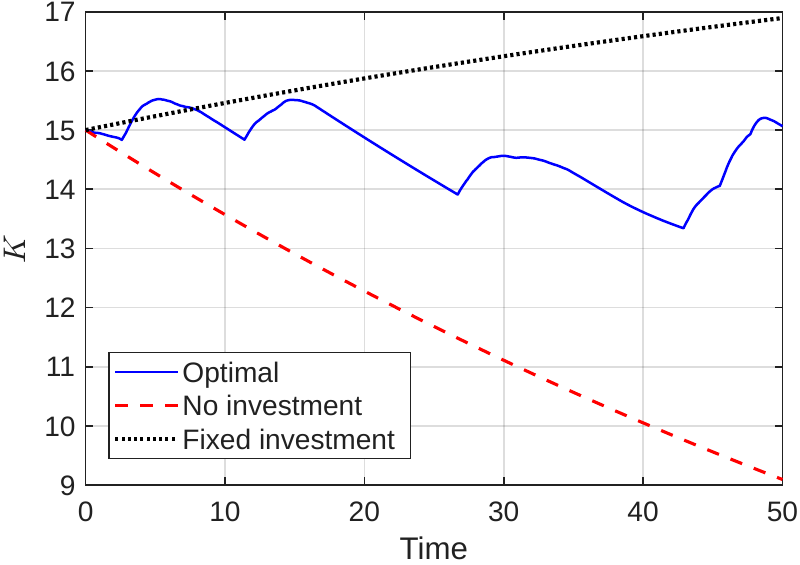}
	\caption{Sample paths of $K$}
	\label{fig:sample K}
\end{figure}

\begin{figure}
	\centering
	\includegraphics[bb=0 0 384 271, width=0.75\textwidth]{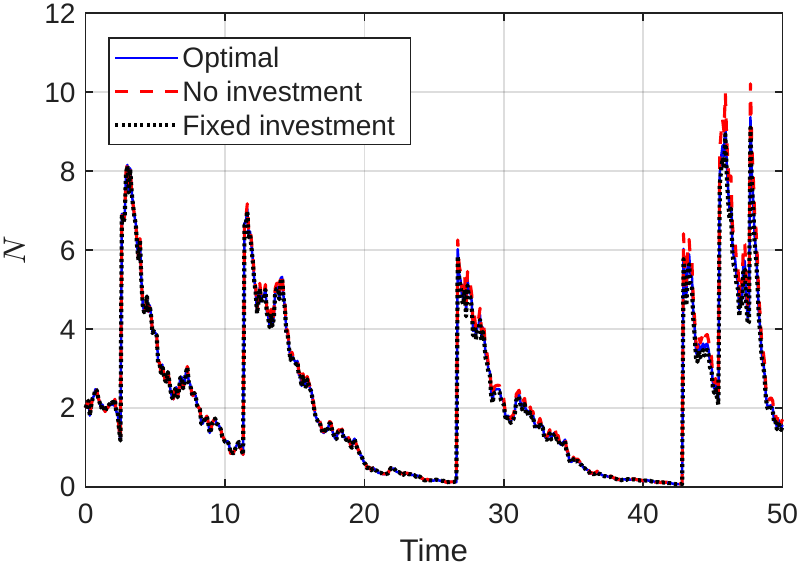}
	\caption{Sample paths of $N$}
	\label{fig:sample N}
\end{figure}

To explore how the investment rules influence the dynamics of the system, we compare the sample paths derived from a single simulation run. 
Figures \ref{fig:sample x}, \ref{fig:sample K}, \ref{fig:sample N} illustrate the trajectories of DRR investment $x$, preparedness capital $K$, and the disaster damage level $N$, respectively, across the three policies.

Under Optimal, investment increases as the damage state deteriorates and declines as the system recovers, with some intervals of zero investment. 
Fixed Investment, by contrast, maintains a constant investment rate regardless of the state of the system.

Preparedness capital gradually increases under Fixed Investment, whereas it rises and falls under Optimal according to the balance between capital accumulation through investment and depreciation. 
In the No Investment scenario, the initial capital stock declines monotonically through depreciation. 
At various points in this scenario, the capital stock under Optimal is observed to be lower than that under Fixed Investment. This trend aligns with the optimization objective, which aims to minimize total discounted losses, encompassing both investment expenditures and disaster-related damages. 
Maintaining a larger preparedness capital stock therefore does not necessarily yield a lower total loss.

The level of disaster damage $N$ experiences jumps at the same arrival times for all three policies.  
However, differences in preparedness capital generate different jump magnitudes through the mitigation function $h(K_t)$. 
Thus, even under identical realizations of the underlying stochastic processes, the damage trajectories differ across the different policies.

\begin{table}[htbp]
    \centering
    \caption{Comparison of investment policies under the benchmark setting}
    \label{tab:benchmark_policy_comparison}
    \small
    \setlength{\tabcolsep}{5pt}
    \renewcommand{\arraystretch}{1.15}

	\begin{tabular}{p{0.51\linewidth}rrr}
		\hline
		Indicator
		& Optimal
		& \begin{tabular}{c}
			No\\Investment
		\end{tabular}
		& \begin{tabular}{c}
			Fixed\\Investment
		\end{tabular} \\
		\hline

		\multicolumn{4}{l}{
			\textit{Expected losses and expenditure}
		} \\
		Expected total loss, $\mathbb{E}[Q^\pi]$
		& 54.55 & 56.20 & 54.86 \\
		Expected 50-year loss, $\mathbb{E}[L_{50}^\pi]$
		& 46.77 & 47.12 & 47.12 \\
		Expected 50-year investment expenditure
		& 4.44 & 0.00 & 4.38 \\
		Expected 50-year disaster damage
		& 42.32 & 47.12 & 42.74 \\
		\hline

		\multicolumn{4}{l}{
			\textit{Distribution of 50-year discounted losses}
		} \\
		Median (P50)
		& 44.11 & 43.56 & 44.14 \\
		P25
		& 33.18 & 32.20 & 33.53 \\
		P75
		& 57.62 & 59.21 & 57.60 \\
		P90
		& 72.05 & 75.24 & 72.54 \\
		P95
		& 82.85 & 86.85 & 84.02 \\
		\hline

		\multicolumn{4}{l}{
			\textit{Upper-tail indicators of 50-year discounted losses}
		} \\
		ES95
		& 98.04 & 103.66 & 100.14 \\
		Tail ratio (P95/Median)
		& 1.88 & 1.99 & 1.90 \\
		\hline
	\end{tabular}
\end{table}

Table \ref{tab:benchmark_policy_comparison} summarizes the simulation results for the three investment policies: Optimal, No Investment, and Fixed Investment. 
It reports expected discounted losses, investment expenditure, and disaster damage, together with indicators describing the distribution and upper-tail risk of 50-year discounted losses.

All indicators are calculated from 500 Monte Carlo runs per policy. 
Within each run, the three policies share the same realizations of the Wiener and Poisson processes.
All loss and expenditure measures are discounted to time zero.
The 50-year loss, $L_{50}^\pi$, includes investment expenditure and disaster damage incurred during the simulation horizon. 
Expected total loss is the sample mean of $Q^\pi=L_{50}^\pi+e^{-\rho T}V^*(K_T^\pi,N_T^\pi)$, where $T=50$. 
The terminal term represents expected continuation losses under the optimal policy after year 50 in all three cases. 
Only expected total loss includes this terminal term. 
The separately reported investment expenditure and disaster damage cover the first 50 years.
Median, percentiles, ES95, and the tail ratio are calculated from $L_{50}^\pi$, excluding the terminal term. 
P$p$ denotes the $p$th percentile.
ES95 is the arithmetic mean of observations greater than or equal to P95.
In the present samples, this mean is calculated from 25 observations for each policy.
The tail ratio is P95 divided by the median and is dimensionless.
Minor discrepancies between totals and their components arise from rounding.

Table \ref{tab:benchmark_policy_comparison} first illustrates the effects of DRR investment by comparing Optimal with No Investment. 
Optimal reduces expected total discounted losses, including the terminal continuation value, from 56.20 to 54.55, a reduction of approximately 2.9\%. 
Over the first 50 years, expected discounted disaster damage falls from 47.12 to 42.32, approximately 10.2\%, while discounted investment expenditure is 4.44. 
After accounting for this expenditure, expected 50-year losses decrease only modestly, from 47.12 to 46.77. The expected discounted continuation losses are also lower under Optimal, at 7.78 compared with 9.08 under No Investment. 
Thus, accounting for the consequences of the terminal state increases the measured advantage of DRR investment.

The advantages of investment are clearly evident in the upper tail of the 50-year loss distribution. 
When comparing the effects of No Investment to Optimal, we observe that the P95 loss decreases from 86.85 to 82.85, while the Expected Shortfall at the 95th percentile (ES95) drops from 103.66 to 98.04. 
This reflects a reduction of approximately 5.4\% in average losses within the upper 5\% of outcomes. 
These findings suggest that, although the mean reduction in 50-year losses is modest, there is a decrease in losses during adverse scenarios.

Comparison with Fixed Investment assesses the effects of adjusting investment to the state of the system. 
The fixed rule matches the average undiscounted investment expenditure under Optimal over the simulation horizon. 
Optimal nevertheless produces slightly lower expected total losses (54.55 vs. 54.86), while the median 50-year losses are almost identical (44.11 vs. 44.14). 
Discounted investment expenditure is slightly higher under Optimal, at 4.44 compared with 4.38, but this additional expenditure is more than offset by lower expected disaster damage, 42.32 compared with 42.74.

State-dependent investment also improves the upper-tail indicators relative to Fixed Investment. 
P95 decreases from 84.02 to 82.85, while ES95 decreases from 100.14 to 98.04, approximately 2.1\%.
The comparison suggests that adjusting investment to evolving conditions can improve loss outcomes at the same average undiscounted investment expenditure, with larger relative improvements in the tail than in mean 50-year losses.

Taken together, the results highlight the distinction between median and tail outcomes. 
No Investment has the lowest median loss (43.56) but the highest upper-tail losses and tail ratio. 
This pattern is consistent with a trade-off between investment expenditure and protection against severe losses. 
The tail ratio decreases from 1.99 under No Investment to 1.90 under Fixed Investment and 1.88 under Optimal. 
In this benchmark setting, the policy that minimizes expected discounted losses also yields the lowest reported upper-tail losses. 
At the same time, state-dependent investment provides additional improvements over a constant investment rule. 
These findings illustrate why DRR policies should be evaluated using both expected losses and the upper tail of the loss distribution.

\clearpage
\subsection{Numerical experiments on model mechanisms}
\label{Simulation}

We consider several alternative cases in which disaster characteristics and the external conditions of the economic system vary to examine their effects on optimal policies and economic dynamics.

This subsection examines the mechanisms of the model through five numerical experiments: (i) increased recurrent uncertainty, represented by a higher diffusion volatility $\beta$; (ii) exposure growth, represented by a larger exposure scale, $\phi$; (iii) greater catastrophic severity, represented by a higher elasticity of economic damage with respect to the damage state, $\eta_N$; (iv) changes in rare-event frequency, represented by variations in the Poisson intensity, $\mu$; and (v) more persistent disaster damage, represented by a lower recovery rate, $\delta_N$. 
Each scenario changes one parameter while holding the remaining parameters at their benchmark values. 
Comparing these scenarios with the benchmark allows us to identify how each mechanism shapes optimal investment, preparedness capital accumulation, and economic loss outcomes.

The recurrent uncertainty scenario considers an environment in which disaster conditions fluctuate more strongly over time. 
A higher $\beta$ increases the variability of the damage state without directly changing its drift or the arrival frequency of catastrophic events. Socioeconomically, this scenario represents greater uncertainty surrounding recurrent hazards, such as stronger fluctuations in rainfall and flood conditions. 
The experiment examines how investment responds to a less predictable damage environment and whether greater variability affects typical losses and upper-tail losses differently.

The exposure growth scenario represents an expansion of economic activity and assets exposed to disaster hazards, as may accompany urbanization or development in hazard-prone areas. 
A higher $\phi$ increases the upward jump in the damage state caused by a catastrophic event for a given stock of preparedness capital. 
The comparison therefore examines whether existing preparedness remains adequate when a larger economy faces the same event frequency. 
It also clarifies how optimal investment and subsequent capital accumulation adjust to the greater potential damage associated with expanded exposure.

The catastrophic severity scenario examines the economic consequences of severe damage through a higher $\eta_N$. 
This parameter governs the elasticity of the economic damage flow with respect to the damage state. 
When the damage function is convex, it captures the disproportionate economic burden that can arise from simultaneous disruptions to transport, utilities, production, and other essential functions. 
The experiment thus focuses on how the economic consequences of a damaged state influence preparedness decisions. 
Although this mechanism is particularly relevant after catastrophic jumps, it also affects losses associated with recurrent fluctuations.

The rare-event frequency scenario changes how often catastrophic events are expected to occur while retaining the same relationship between preparedness capital and the damage increment per event. 
Variations in $\mu$ can represent different hazard environments or changes in the occurrence rate of extreme events. 
This comparison examines how event frequency influences investment incentives and the maintenance of preparedness capital between disasters. 
It also helps distinguish the effects of more frequent catastrophic events from those of larger damage increments or greater economic sensitivity to damage.

The damage persistence scenario represents slower recovery from disaster impacts. 
A lower $\delta_N$ reduces the rate at which the damage state declines between shocks, allowing disruptions and their associated economic losses to persist for longer. 
Socioeconomic interpretations include longer infrastructure outages, delayed production restoration, and limited recovery capacity. 
This experiment examines how a greater expected burden from persistent damage changes optimal preparedness investment. 
Because the recovery parameter acts on the damage state regardless of its origin, it affects the persistence of damage generated by both recurrent fluctuations and catastrophic events.

Together, these comparisons clarify how variability, exposure, economic damage sensitivity, event frequency, and recovery interact with endogenous preparedness. 
We examine both investment and capital trajectories, together with the level and distribution of losses, to determine whether these mechanisms produce different responses in typical and upper-tail outcomes. 
The purpose is to explain why distinct disaster risk environments may require different patterns of investment and preparedness accumulation, even when all policies minimize expected discounted economic losses.
Table \ref{tab:parameter variations} shows the parameter changes set for each scenario.

\begin{table}
	\centering
	\caption{Parameter variations for numerical experiments}
	\label{tab:parameter variations}
	\begin{tabular}{cccc}
		\hline
		scenario	&	parameter	&	benchmark	&	variation	\\ \hline
		recurrent uncertainty	&	$\beta$	&	0.28	&	0.336	\\
		exposure growth	&	$\phi$	&	10	&	15	\\
		catastrophic severity	&	$\eta_N$	&	1.0	&	1.05	\\
		rare-event frequency	&	$\mu$	&	0.15	&	0.18	\\
		more persistent damage	&	$\delta_N$	&	0.2	&	0.16	\\
		\hline
	\end{tabular}
\end{table}

\begin{figure}
	\footnotesize
	\begin{minipage}{0.5\linewidth}
		\centering
		\includegraphics[bb=0 0 350 237, width=0.975\textwidth]{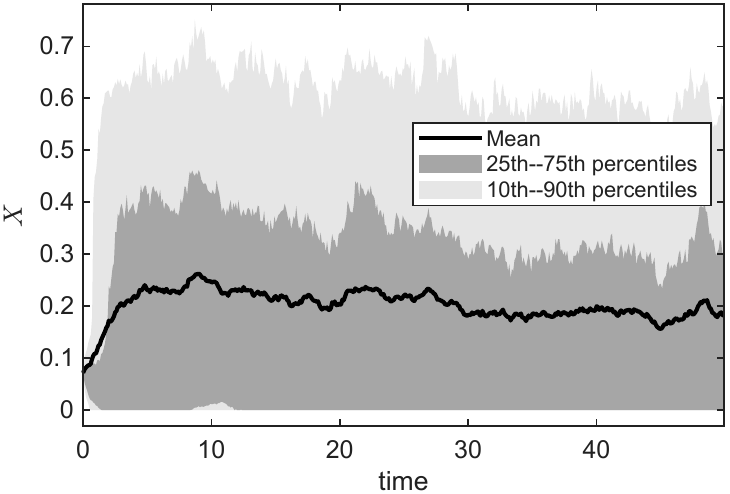}
		benchmark
	\end{minipage}
	\begin{minipage}{0.5\linewidth}	
		\centering
		\includegraphics[bb=0 0 350 237, width=0.975\textwidth]{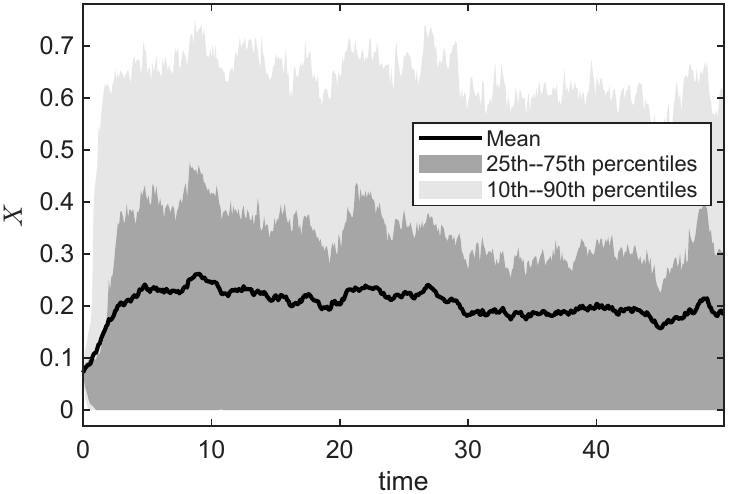}
		recurrent uncertainty
	\end{minipage}
	\vspace{5mm}
	\\
	\begin{minipage}{0.5\linewidth}	
		\centering
		\includegraphics[bb=0 0 350 237, width=0.975\textwidth]{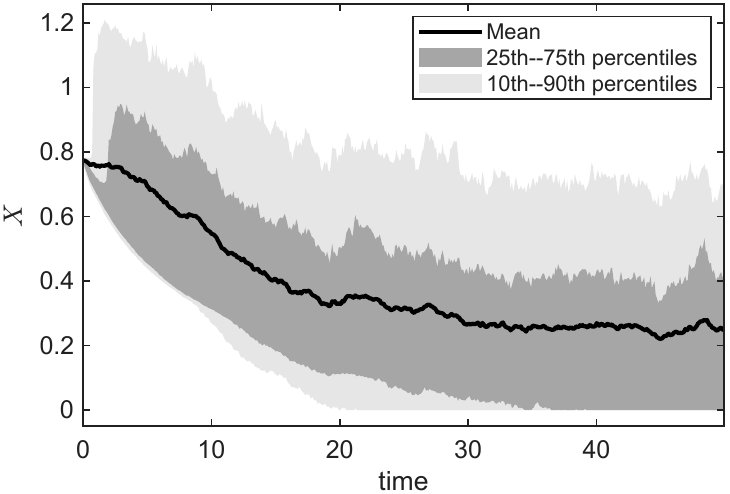}
		exposure growth
	\end{minipage}
	\begin{minipage}{0.5\linewidth}
		\centering
		\includegraphics[bb=0 0 350 237, width=0.975\textwidth]{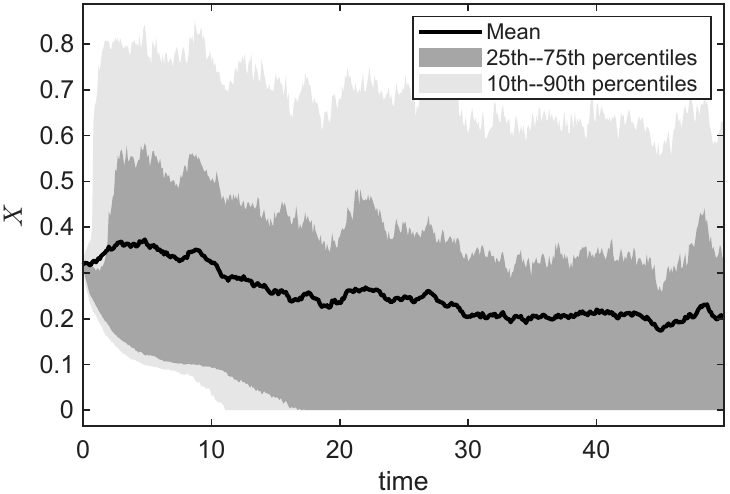}
		catastrophic severity
	\end{minipage}
	\vspace{5mm}
	\\
	\begin{minipage}{0.5\linewidth}	
		\centering
		\includegraphics[bb=0 0 350 237, width=0.975\textwidth]{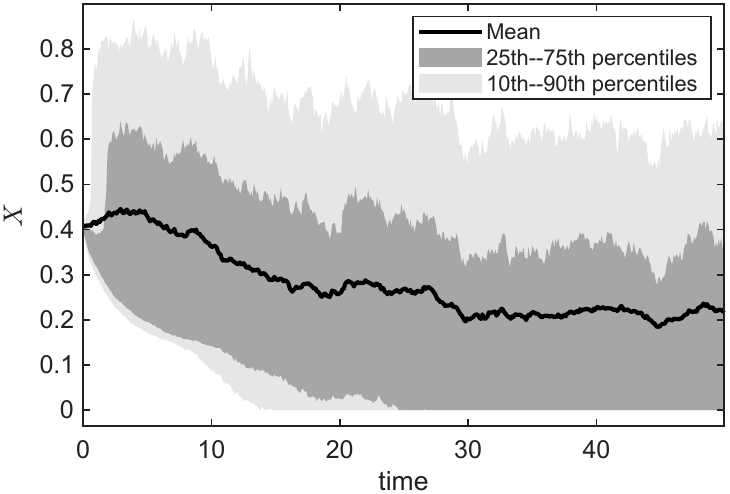}
		rare-event frequency
	\end{minipage}
	\begin{minipage}{0.5\linewidth}
		\centering
		\includegraphics[bb=0 0 350 237, width=0.975\textwidth]{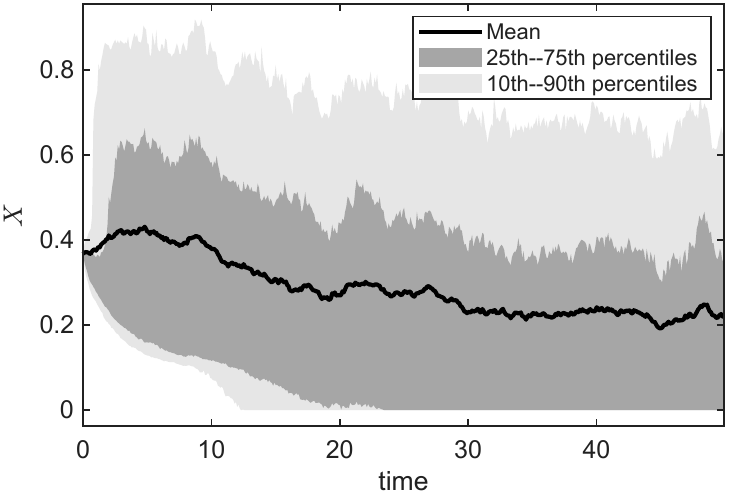}
		more persistent damage
	\end{minipage}
	\caption{Dynamics of $x^{\ast}$}
	\label{fig:scenario dynamics x}
\end{figure}

\begin{figure}
	\footnotesize
	\begin{minipage}{0.5\linewidth}
		\centering
		\includegraphics[bb=0 0 350 237, width=0.975\textwidth]{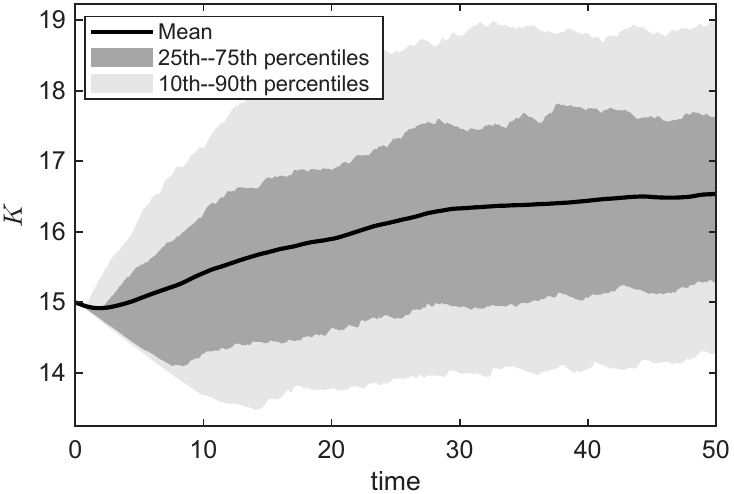}
		benchmark
	\end{minipage}
	\begin{minipage}{0.5\linewidth}	
		\centering
		\includegraphics[bb=0 0 350 237, width=0.975\textwidth]{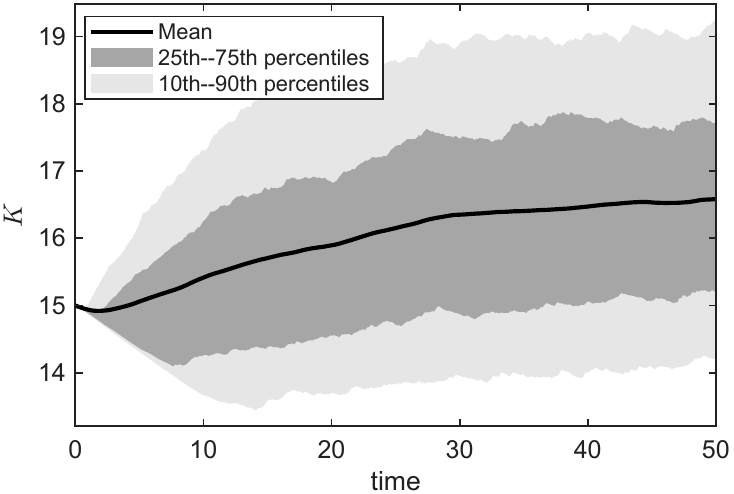}
		recurrent uncertainty
	\end{minipage}
	\vspace{5mm}
	\\
	\begin{minipage}{0.5\linewidth}	
		\centering
		\includegraphics[bb=0 0 350 237, width=0.975\textwidth]{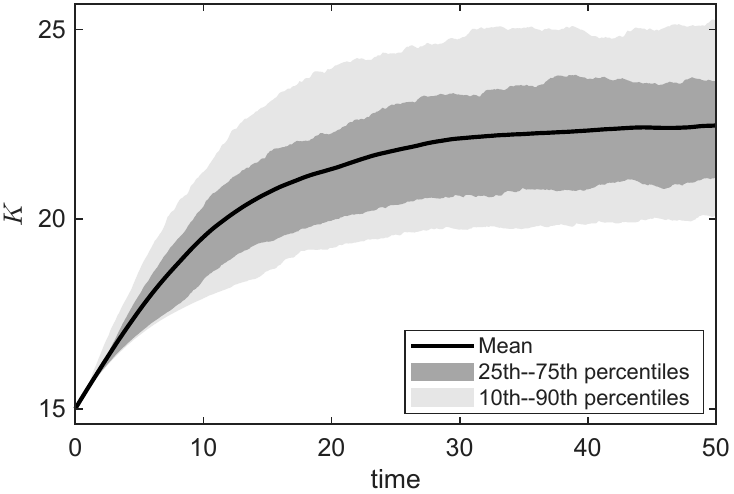}
		exposure growth
	\end{minipage}
	\begin{minipage}{0.5\linewidth}
		\centering
		\includegraphics[bb=0 0 350 237, width=0.975\textwidth]{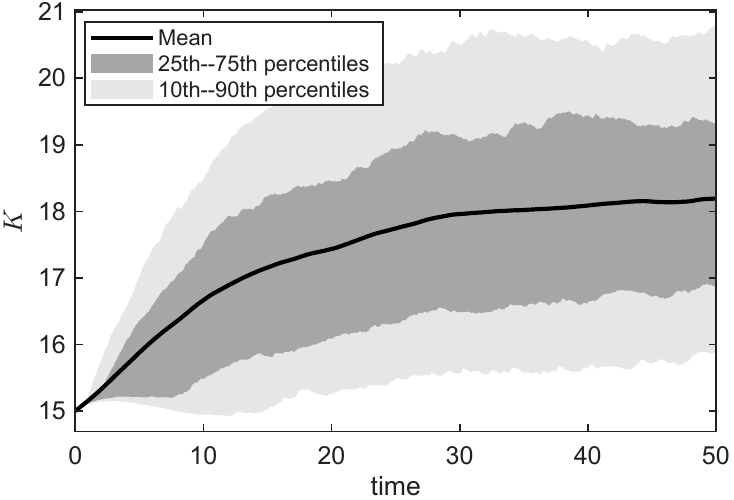}
		catastrophic severity
	\end{minipage}
	\vspace{5mm}
	\\
	\begin{minipage}{0.5\linewidth}	
		\centering
		\includegraphics[bb=0 0 350 237, width=0.975\textwidth]{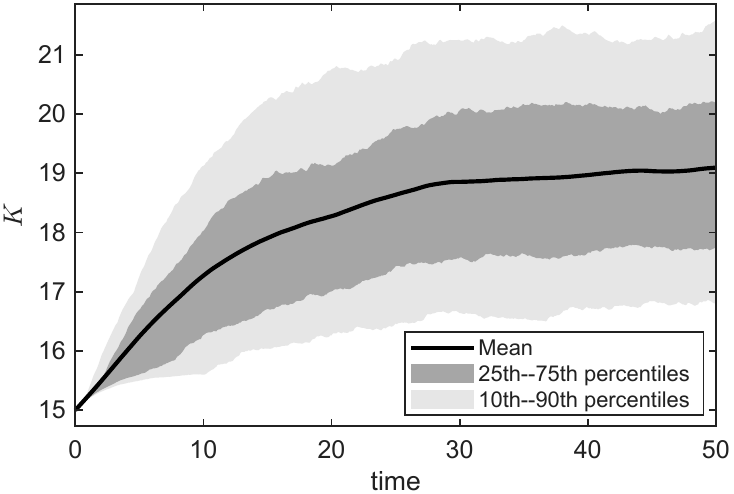}
		rare-event frequency
	\end{minipage}
	\begin{minipage}{0.5\linewidth}
		\centering
		\includegraphics[bb=0 0 350 237, width=0.975\textwidth]{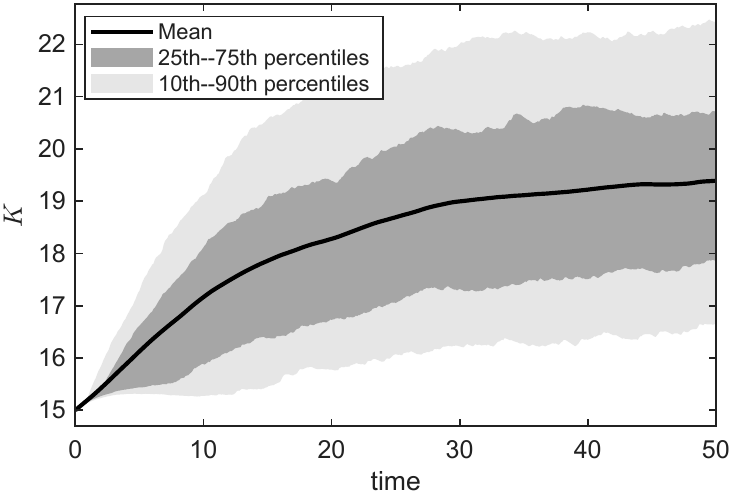}
		more persistent damage
	\end{minipage}
	\caption{Dynamics of $K$}
	\label{fig:scenario dynamics K}
\end{figure}

\begin{figure}
	\footnotesize
	\begin{minipage}{0.5\linewidth}
		\centering
		\includegraphics[bb=0 0 350 237, width=0.975\textwidth]{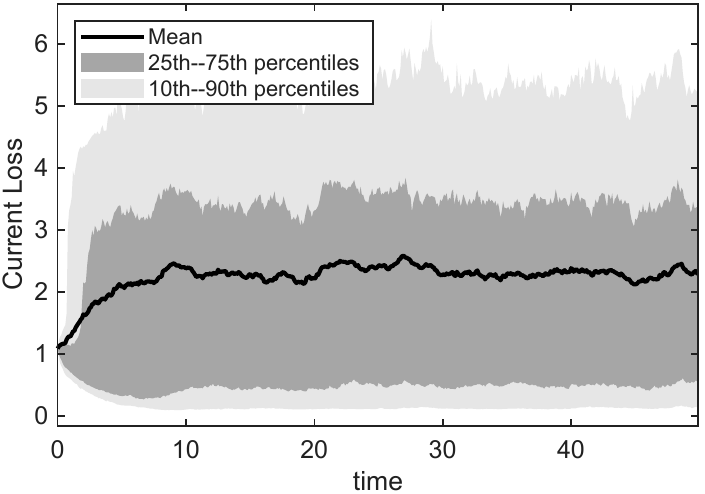}
		benchmark
	\end{minipage}
	\begin{minipage}{0.5\linewidth}	
		\centering
		\includegraphics[bb=0 0 350 237, width=0.975\textwidth]{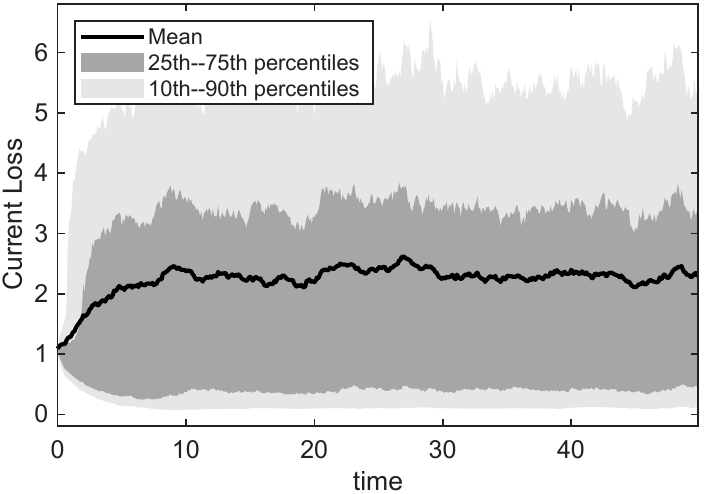}
		recurrent uncertainty
	\end{minipage}
	\vspace{5mm}
	\\
	\begin{minipage}{0.5\linewidth}	
		\centering
		\includegraphics[bb=0 0 350 237, width=0.975\textwidth]{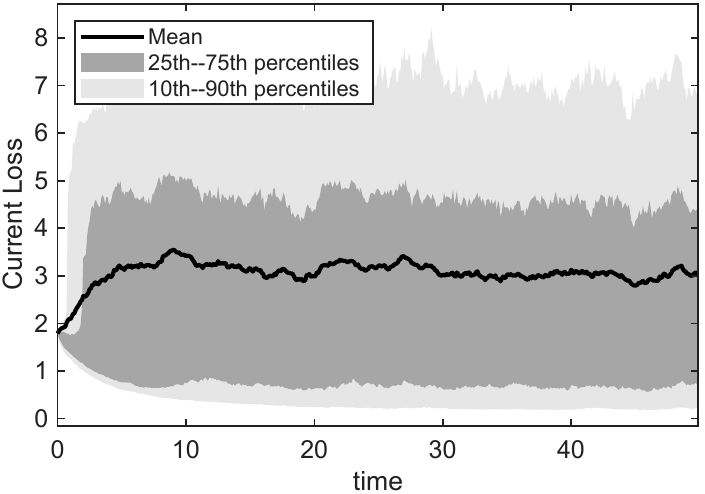}
		exposure growth
	\end{minipage}
	\begin{minipage}{0.5\linewidth}
		\centering
		\includegraphics[bb=0 0 350 237, width=0.975\textwidth]{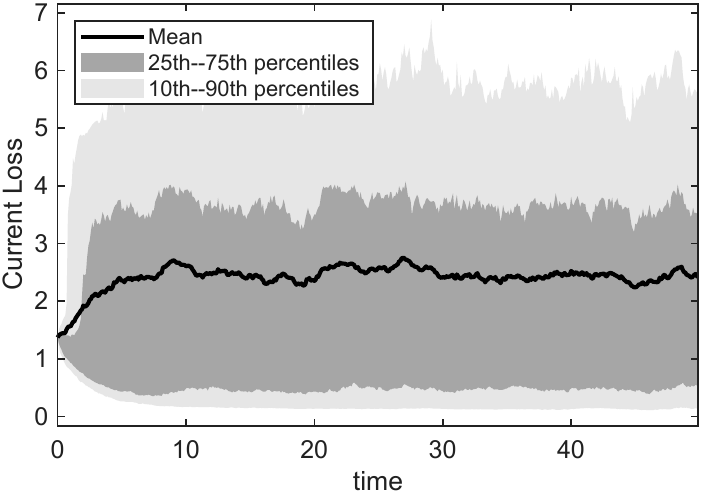}
		catastrophic severity
	\end{minipage}
	\vspace{5mm}
	\\
	\begin{minipage}{0.5\linewidth}	
		\centering
		\includegraphics[bb=0 0 350 237, width=0.975\textwidth]{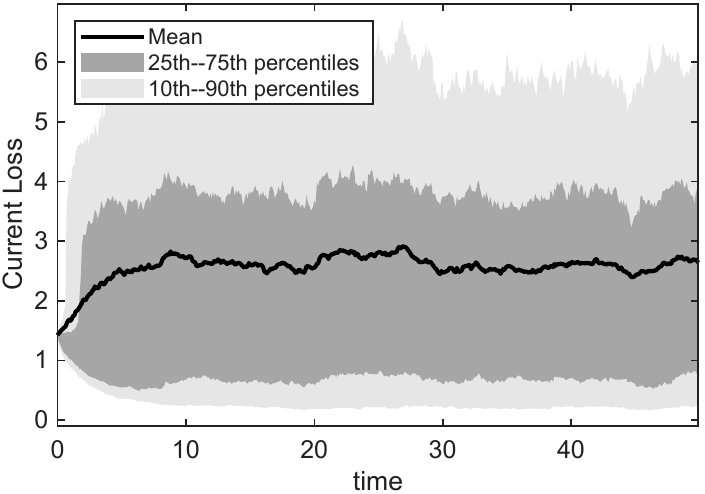}
		rare-event frequency
	\end{minipage}
	\begin{minipage}{0.5\linewidth}
		\centering
		\includegraphics[bb=0 0 350 237, width=0.975\textwidth]{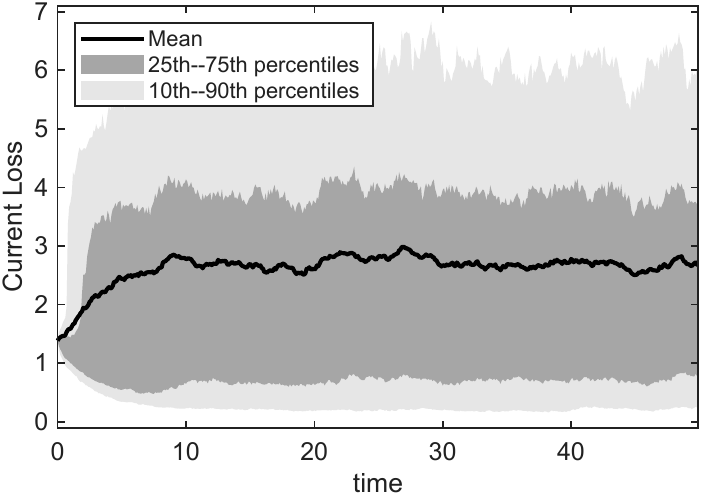}
		more persistent damage
	\end{minipage}
	\caption{Dynamics of current loss}
	\label{fig:scenario dynamics CL}
\end{figure}

Figures \ref{fig:scenario dynamics x}, \ref{fig:scenario dynamics K}, \ref{fig:scenario dynamics CL} illustrate the dynamics of investment $x$, disaster preparedness capital $K$, and current-period economic loss (current loss).
These figures show the mean values of each variable for each period, along with the 25th--75th and 10th--90th percentile ranges, based on 500 runs.

\subsubsection{Benchmark}
Initially, average investment rises, reaching a relatively high level around the 10-year mark before gradually declining. Capital $K$ initially decreases slightly, then rises steadily, ultimately averaging approximately 16.5 after 50 years. This growth from the initial capital of 15 occurs gradually. Meanwhile, the average current loss rises initially before stabilizing at a roughly constant level.

\subsubsection{Recurrent uncertainty}
The recurrent uncertainty scenario sets $\beta$ 20\% higher than the benchmark.
The trends in the average values of $x$ and $K$ are very similar to the benchmark, and current loss levels differ little.
Although the difference appears small in the figure alone, Table \ref{tab:scenario_comparison} reveals distinctive characteristics in the distribution of cumulative losses.
The expected loss over 50 years changes from 46.77 to 46.68, remaining nearly the same.
The median decreases by 1.5\%, from 44.11 to 43.46.
However, P95 increases by 2.8\% from 82.85 to 85.16, and ES95 increases by 4.7\% from 98.04 to 102.61.
In this experiment, greater recurrent uncertainty therefore raises upper-tail losses while leaving expected losses nearly unchanged.

\subsubsection{Exposure growth}
The mean value of $x$ begins at a high point of approximately 0.8, then declines largely. Correspondingly, $K$ accumulates rapidly, reaching an average value of approximately 22.5 after 50 years. Both the initial investment and the final value of $K$ are the highest among the scenarios compared.
Current losses continue to surpass the benchmark, with the cumulative loss metric standing as the highest across all cases. 
Expected loss over 50 years is 65.75, a 40.6\% increase over the benchmark.
The median of 61.80, P95 of 115.55, and the expected shortfall at the 95th percentile of 136.41 represent increases of approximately 39\% to 40\%, respectively.
Expected investment costs are projected to rise by 131.5\%, from 4.44 to 10.28.
Additionally, expected disaster losses are anticipated to increase by 31.0\%, rising from 42.32 to 55.46. 
While the expansion in exposure scale contributes to a substantial rise in capital, it also results in an upward shift of the entire loss distribution.  
Notably, the median and the upper tail both increase by roughly the same percentage, keeping the tail ratio relatively stable with values of 1.88 and 1.87. However, despite the relative tail metrics remaining nearly unchanged, absolute tail losses have markedly increased.

\subsubsection{Catastrophic severity}
In the catastrophic severity scenario, mean investment $x$ starts at approximately 0.3, rises briefly, and then gradually declines toward levels close to the benchmark. Mean preparedness capital $K$ increases from 15 to approximately 18.2 after 50 years, compared with approximately 16.5 in the benchmark. Although capital accumulation is less pronounced than in the exposure growth and rare-event frequency scenarios, the preparedness stock remains above the benchmark throughout the subsequent period. Mean current losses nevertheless generally remain above the benchmark.
The expected investment cost is 6.10, reflecting a 37.4\% increase compared to the benchmark.
The expected disaster loss is 44.92, marking a 6.1\% rise, while the expected loss over 50 years is 51.02, indicating a 9.1\% increase.
The median increases by 9.3\%, the P95 by 10.6\%, and the ES95 by 11.1\%.
In this case, both median losses and tail losses increase, with the rate of increase being slightly higher at the upper end. The tail ratio is 1.90, representing a slight increase from the benchmark.

\subsubsection{Rare-event frequency}
In this scenario, increasing $\mu$ by 20\% represents an upward shift in the frequency of rare disasters.
The mean $x$ starts at approximately 0.4, rises briefly, and then declines as capital accumulates. Unlike the benchmark’s path, which starts with a low initial investment and gradually increases, this scenario involves strengthening disaster preparedness capital early on in response to the higher disaster frequency. Average capital increases from 15 to approximately 19, exceeding the benchmark’s level of approximately 16.5 after 50 years. Average current losses also remain higher than the benchmark.
Expected investment costs over 50 years increase by 56.3\%, while expected disaster losses also rise by 11.5\%; combined, expected losses increase by 15.7\%. Even with capital buildup achieved through the optimal policy, the economic burden associated with the increased frequency of disasters remains.
The median cumulative loss rises by 18.7\%, with the P95 up 12.8\% and the ES95 up 7.1\%. As a result, the tail ratio decreases from 1.88 to 1.78. This increase in frequency has led to a notable rise in central losses, while the ratio of tail losses to the median has diminished. However, since P95 and ES95 themselves have increased, this cannot be interpreted as a reduction in tail losses.

\subsubsection{More persistent damage}
In this scenario, $\delta_N$ falls by 20\%, from 0.20 to 0.16. Average investment rises from approximately 0.36 in the early stages and then gradually declines. Corresponding to the higher initial investment compared to the benchmark, $K$ accumulates, and the average capital after 50 years reaches approximately 19.4, exceeding the benchmark’s approximately 16.5. Meanwhile, average current-period losses remain higher than the benchmark, and even as $K$ accumulates, the economic burden from delayed recovery persists.
The expected investment cost over 50 years increases from 4.44 to 7.03, and the expected disaster damage increases from 42.32 to 47.82. The combined expected loss increases by 17.3\%, from 46.77 to 54.85. This indicates that damage persists longer due to the slower recovery rate, and that cumulative losses increase even when investment rises under the optimal policy.
The median cumulative loss increases by 16.8\% from 44.11 to 51.51, while P95 increases by 21.4\% from 82.85 to 100.62 and ES95 increases by 19.1\% from 98.04 to 116.79. Since the upper-tail increase exceeds the median, the tail ratio also rises from 1.88 to 1.95. Therefore, a delayed recovery not only pushes up typical loss levels but also strongly affects the cumulative losses of trials in which serious damage persists for an extended period.

\clearpage

\clearpage

\begin{table}[htbp]
    \centering
    \caption{Comparison of optimal-policy outcomes across disaster risk scenarios}
    \label{tab:scenario_comparison}
    \small
    \setlength{\tabcolsep}{4pt}
    \renewcommand{\arraystretch}{1.15}

    \resizebox{\linewidth}{!}{%
    \begin{tabular}{lrrrrrr}
        \hline
        Indicator
        & Benchmark
        & \begin{tabular}{c}Recurrent\\uncertainty\end{tabular}
        & \begin{tabular}{c}Exposure\\growth\end{tabular}
        & \begin{tabular}{c}Catastrophic\\severity\end{tabular}
        & \begin{tabular}{c}Rare-event\\frequency\end{tabular}
        & \begin{tabular}{c}More persistent\\damage\end{tabular} \\
        \hline

        \multicolumn{7}{l}{
            \textit{Expected losses and expenditure}
        } \\
        Expected total loss, $\mathbb{E}[Q^s]$
        & 54.55 & 54.45 & 75.86 & 59.22 & 62.86 & 63.85 \\
        Expected 50-year loss, $\mathbb{E}[L_{50}^s]$
        & 46.77 & 46.68 & 65.75 & 51.02 & 54.11 & 54.85 \\
        Expected 50-year investment expenditure
        & 4.44 & 4.48 & 10.28 & 6.10 & 6.94 & 7.03 \\
        Expected 50-year disaster damage
        & 42.32 & 42.20 & 55.46 & 44.92 & 47.17 & 47.82 \\
        \hline

        \multicolumn{7}{l}{
            \textit{Distribution of 50-year discounted losses}
        } \\
        Median (P50)
        & 44.11 & 43.46 & 61.80 & 48.22 & 52.36 & 51.51 \\
        P25
        & 33.18 & 32.39 & 47.62 & 36.04 & 39.54 & 38.65 \\
        P75
        & 57.62 & 58.32 & 80.14 & 62.69 & 66.40 & 67.86 \\
        P90
        & 72.05 & 73.21 & 99.83 & 79.01 & 80.82 & 84.88 \\
        P95
        & 82.85 & 85.16 & 115.55 & 91.66 & 93.42 & 100.62 \\
        \hline

        \multicolumn{7}{l}{
            \textit{Upper-tail indicators of 50-year discounted losses}
        } \\
        ES95
        & 98.04 & 102.61 & 136.41 & 108.93 & 104.99 & 116.79 \\
        Tail ratio (P95/Median)
        & 1.88 & 1.96 & 1.87 & 1.90 & 1.78 & 1.95 \\
        \hline
    \end{tabular}%
    }

    \vspace{0.5em}
    \begin{minipage}{\linewidth}
        \footnotesize
        \textit{Notes:}
        $s$ denotes a scenario. Each scenario is evaluated
        under its corresponding optimal policy using 500 simulation runs.
        All loss and expenditure measures are discounted to time zero.
        Expected total loss includes the terminal continuation term,
        following the convention in Section~4.2.
        Median, percentiles, ES95, and the tail ratio refer to
        50-year discounted losses, excluding the terminal term.
        ES95 summarizes average losses in the upper 5\% tail.
        The tail ratio is P95 divided by the median and is dimensionless.
        Minor discrepancies between totals and their components
        arise from rounding.
    \end{minipage}
\end{table}



\subsection{Implications for DRR investment policy}
\label{discussion}
The scenario analysis indicates that DRR investment planning should not only consider the magnitude of disaster risk but also the underlying mechanisms that contribute to it. Various factors identified in our analysis—including damage volatility, the scale of exposure, the sensitivity of economic losses to damage, the frequency of rare disasters, and the speed of recovery—each influence the optimal investment levels and the distribution of losses in distinct ways.
Consequently, instead of applying a uniform investment increase in response to shifts in the risk environment, planning should comprehensively assess investment timing, the accumulation of preparedness capital, and residual losses. Below, we outline the policy implications derived from this analysis.

First, annual investment expenditure alone cannot assess long-term preparedness. Under exposure growth, increased rare-event frequency, and more persistent damage, relatively high initial investment builds preparedness capital, after which average investment declines. This decline does not imply a reduction in preparedness capacity. Investment planning should therefore consider the existing capital stock, its depreciation, and its adequacy under future risk conditions alongside current expenditure. Moreover, the average investment trajectory aggregates numerous state-dependent decisions and should not be adopted directly as a fixed budget allocation rule.

Second, evaluating an expected-loss-minimizing policy requires indicators of the loss distribution. Increased recurrent uncertainty raises P95 and ES95 even though expected losses remain almost unchanged. By contrast, increased rare-event frequency raises the median proportionally more than upper-tail losses, reducing the tail ratio. Exposure growth also substantially increases absolute upper-tail losses while leaving the tail ratio nearly unchanged. These results show that expected losses or relative tail indicators alone cannot fully characterize preparedness needs for large losses. Policy evaluation should incorporate a clear distinction between expected and median losses, alongside the assessment of absolute upper-tail losses. 

Third, rare-event frequency and damage persistence have different policy implications. The increased-frequency and slower-recovery scenarios produce relatively similar expected investment expenditures and expected losses, but P95 and ES95 are higher under slower recovery. This finding suggests that DRR planning should consider the duration of functional disruptions and economic stagnation alongside event counts and average damage. Preparedness investment should therefore be considered together with restoration and recovery capacity that limits prolonged damage. However, recovery speed is exogenous in this model, so the analysis does not evaluate the optimal allocation between recovery measures and preparedness investment or their comparative cost-effectiveness.

These findings highlight the need to combine capital accumulation and maintenance, state-dependent investment adjustment, and assessment of residual upper-tail losses in long-term DRR planning. The tail indicators in this study provide ex post evaluations of losses generated under an expected-loss-minimizing policy; they do not represent outcomes from directly minimizing tail losses. Furthermore, because the parameter changes differ across scenarios, the numerical results cannot establish a general ranking of the importance of different risk factors. The policy contribution of this analysis lies in providing a framework for evaluating investment trajectories and loss distributions based on the structure of disaster risk, rather than recommending specific investment amounts.

\section{Conclusion}
\label{conclusion}

This study developed a stochastic dynamic control framework for long-term disaster risk reduction investment under recurrent and catastrophic risks. By jointly representing continuous disaster fluctuations and rare catastrophic shocks, and by treating preparedness capital as an endogenous stock accumulated through investment, the model links current DRR decisions to both the prevailing damage state and previously accumulated preparedness capacity. This formulation provides a unified framework for examining how disaster preparedness should adjust dynamically rather than be represented as a fixed expenditure rule or a one-time project decision.

The numerical analysis yields a broader conclusion that extends beyond the outcomes of individual scenarios. Effective DRR policy depends not only on the overall magnitude of disaster risk, but also on the mechanism through which that risk evolves. Changes in uncertainty, exposure, disaster consequences, event occurrence, and recovery conditions can affect investment requirements, preparedness accumulation, and the distribution of economic losses in different ways. Consequently, neither current investment expenditure nor a single measure of expected loss can characterize preparedness needs. Long-term evaluation should instead consider the interaction among state-dependent investment, accumulated preparedness capital, recovery dynamics, and the distribution of residual losses. The benchmark policy comparison further supports this interpretation by showing that the value of DRR policy derives not only from undertaking preparedness investment, but also from adjusting that investment to evolving economic and disaster conditions. 

These findings suggest a shift in DRR investment planning. Preparedness should be viewed as a dynamically managed stock rather than an annual expenditure target, and investment policy should be evaluated alongside other dimensions of risk management, including exposure and recovery capacity. Measures of upper-tail loss can complement expected-loss evaluation by revealing residual risks that are not apparent from central measures alone. However, these indicators are ex post diagnostics under the expected-loss-minimizing policy developed in this study, not objectives the model directly optimizes. More generally, the contribution of the framework lies in identifying how the structure of disaster risk shapes desirable preparedness trajectories, rather than in prescribing a universally optimal level of investment.

The analysis remains stylized. The benchmark parameters represent a hypothetical disaster-risk environment and are not fully calibrated or validated for a specific region. The numerical comparisons therefore illustrate model mechanisms rather than establish a general ranking of risk factors or provide direct investment prescriptions. Future research should empirically calibrate and validate the framework using observed disaster losses, hazard occurrences, recovery processes, exposed assets, and DRR expenditures. Further extensions could endogenize recovery investment and fiscal constraints, and incorporate spatial heterogeneity and interregional risk dependence. Such developments would strengthen the framework as a practical tool for long-term DRR planning.

\clearpage
\appendix
\section{Derivation of HJB equation}
\label{app 1}

Along the optimal path under an infinite horizon, the principle of optimality yields:

\begin{equation}
	\label{oprimality principle}
	\begin{split}
		\rho V(K,N) &= \\
		\min_{\{\boldsymbol{x} \} } &\left[
			\left(x + N^{\eta_N} K^{-\eta_K} \right) 
			+ \frac{E \left(dV(K,N) \right) }{dt}
		\right]
	\end{split}
\end{equation}

Applying the generalized Ito formula for jump-diffusion processes \cite{dixit1994} to the second term on the right-hand side yields the Hamilton-Jacobi-Bellman (HJB) equation. As the first step in this procedure, taking the expected value of the time derivative of $V$ that incorporates both the jump process and the diffusion process for the second term on the right-hand side of (\ref{oprimality principle}) gives:

\begin{equation}
	\label{OBL Poisson expected value of value function differential}
	\begin{split}
		E\left[dV\right] =& \biggl\{
			\left( x - \frac{\alpha}{2} {x^2} -\delta_K K \right)
			\frac{\partial V}{\partial K} - \delta_N N 
			\frac{\partial V}{\partial N}
		\biggr\} dt \\
		&+ \mu \left\lbrace
			V\left(
				K,N + \frac{\phi }{K^{\gamma}}
			\right)
			-V\left(K,N\right) 
		\right\rbrace
		dt\\
		&+ \left(
			\frac{\beta^2 N^2}{2} \frac{\partial^2 V}{\partial N^2}
		\right) dt \\
	\end{split}
\end{equation}

\section{Collocation method for solving HJB equation}
\label{app 2}

\subsection{Outline of the numerical method}

The HJB equation of the stochastic control problem this study constructs is a functional estimation problem that generally admits no closed-form analytical solution, and analyzing its properties requires numerical simulation over a specified range of state variables. Here we summarize the outline of the numerical procedure.
First, we approximate the value function as $V(K,N)\approx \Phi(K,N) \boldsymbol{c}$, where $\boldsymbol{c}$ denotes a coefficient vector whose dimension equals the number of terms in the tensor product of the basis functions for $\boldsymbol{K}$ and $\boldsymbol{N}$, respectively. We then define intervals for the state variables $K$ and $N$ over the range of interest and approximate the continuous space with a discrete grid. Specifically, we define discrete points ${K_0,K_1,\cdots,K_{mK}}$ over the interval $K_l \le K \le K_u$, and discrete points ${N_0,N_1,\cdots,N_{mN}}$ over the interval $N_l \le N \le N_u$.
After completing these preprocessing steps, we express the HJB equation in terms of the polynomial approximation of the value function $V$ and derive the polynomials for the first- and second-order partial derivatives of $V$. We then update the coefficient vector so that the resulting functional satisfies the HJB equation at all discrete points.

\subsection{Details of the computational procedure}

The detailed computational procedure is as follows.
\begin{enumerate} 
	\item Prepare the polynomial function $\Phi(K,N)\boldsymbol{c}$ and assign appropriate initial values to the coefficient vector $\boldsymbol{c}$. In this study, we employ Chebyshev polynomials of the first kind, in which the recurrence relation defines the basis of degree $n$ in variable $x$: 
	\begin{align}
		T_0(x) & =1 \notag \\
		T_1(x) & =x \notag \\
		T_{n+1} (x) &= 2x T_{n}(x) -T_{n-1}(x) \notag
	\end{align} 
	Since the state space is two-dimensional, comprising $K$ and $N$, the polynomial approximation of the value function is defined as: 
	\begin{align}
		V\left(K_r,N_s \right) & \approx \Phi \left(K_r,N_s \right) \boldsymbol{c} \notag \\
		& = \sum_{j_1=0}^{m_1} \sum_{j_2=0}^{m_2} c_{j_1 j_2} 
			T_{j_1} \left(K_r \right)
			T_{j_2} \left(N_s \right) \notag \\
		& \equiv \left[T_{j_2} \otimes T_{j_1} \right] \boldsymbol{c}
	\end{align} 
	The degrees $m_1=3$ and $m_2=3$ were selected through iterative feedback from the convergence results.
	\item At each node $(K_i, N_j)$ of the state variables, obtain the optimal control $x^{\ast}$ of the HJB equation from equation (\ref{derivation optim x}). Under the present specification, the explicit solution $x_{i,j}^{\ast}=x^{\ast}(K_i,N_j;\boldsymbol{c})$ is derived as:
	\begin{equation*}
		\hat{x}_{i,j}=
			\frac{1}{\alpha} \left( 1 + \frac{1}{V_K(K_i,N_j)}\right)
		= \frac{1}{\alpha} \left\lbrace
		1 + \frac{1}{
			\frac{\partial \Phi\left(K_i, N_j\right) \boldsymbol{c}}{\partial K }
		}
		\right\rbrace
	\end{equation*}
	\begin{equation*}
		x^{*}_{i,j}=
		\max \left\lbrace 
			\hat{x}_{i,j},0
		\right\rbrace
	\end{equation*}

	\item Substituting the above first-order condition into the HJB equation yields:
	\begin{equation}
		\label{procedure HJB}
		\begin{split}
			\rho \Phi(K_i,N_j) \boldsymbol{c} =& 
			x_{i,j}^{\ast} +\frac{ {N_j}^{\eta_N}}{{K_i}^{\eta_K}} \\
			&+ \Biggl\{
				\left( x_{i,j} - \frac{\alpha}{2} ({x_{i,j}}^{\ast})^2 -\delta_K K_i \right) 
			\Biggr. \\
			& \Biggl.
			\frac{\partial \Phi \left(K_i, N_j\right)}{\partial K} \boldsymbol{c}
				- \delta_N N_j 
					\frac{\partial \Phi \left(K_i, N_j\right)}{\partial N} \boldsymbol{c}
			\Biggr\} \\
			&+ \mu \left\{ 
				\Phi \left(K_i, N_j+\phi K_i^{-\gamma}\right) \boldsymbol{c} 
				\right. \\
			& \left. -\Phi \left(K_i, N_j\right) \boldsymbol{c}
				\right\} \\
			& + \left(
				\frac{\beta^2 N_j^2}{2}
				\frac{\partial^2 \Phi \left(K_i, N_j\right)}{\partial N^2}
				\boldsymbol{c}
			\right)
		\end{split}
	\end{equation}
	Introducing the following shorthand notation:
	\begin{equation*}
		\label{procedure f}
		f_{i,j} = x_{i,j}^{\ast} +\frac{ N_j^{\eta_N}}{K_i^{\eta_K}}
	\end{equation*}
	\begin{equation*}
		\label{procedure m}
		m_{i,j} =	\left( 
			x_{i,j}^{\ast} - \frac{\alpha}{2} ({x_{i,j}^{\ast}})^2 -\delta_K K_i 
		\right) 
	\end{equation*}
	the condition that must hold at all nodes $(K_i, N_j)$ for the value function becomes:
	\begin{equation}
		\label{procedure approximate}
		\begin{split}
			\rho \Phi \boldsymbol{c} =&
			f_{i,j} + m_{i,j} \frac{\partial \Phi \left(K_i, N_j\right)}{\partial K} \boldsymbol{c}
			- \delta_N N_j \frac{\partial \Phi \left(K_i, N_j\right)}{\partial N} \boldsymbol{c} \\
			& +\mu \Bigl\{ 
				\Phi \left(K_i, N_j+ \phi K_i^{-\gamma}\right) \boldsymbol{c} 
			\Bigr. \\
			& \Bigl. -\Phi \left(K_i, N_j\right) \boldsymbol{c} \Bigr\} \\
			& + \frac{\beta^2 {N_j}^2}{2} 
			\frac{\partial^2 \Phi \left(K_i, N_j\right)}{\partial N^2}
			\boldsymbol{c}
		\end{split}
	\end{equation}

	\item Rearranging by separating terms involving the coefficient vector $\boldsymbol{c}$ from the remaining terms gives:
	\begin{equation}
		\label{rewrite form}
		\begin{split}
			& \Bigl[
				\rho \Phi - m_{i,j} \frac{\partial \Phi \left(K_i, N_j\right)}{\partial K}
				+ \delta_N N_j \frac{\partial \Phi \left(K_i, N_j\right)}{\partial N}
			\Bigr.\\
			& -\mu \left\lbrace 
				\Phi \left(K_i, N_j+\phi K_i^{-\gamma}\right) -\Phi \left(K_i, N_j\right) 
			\right\rbrace \\
			& \left.- \frac{\beta^2 {N_j}^2}{2} \frac{\partial^2 \Phi \left(K_i, N_j\right)}{\partial N^2}
			\right] \boldsymbol{c} - f_{i,j} =0
		\end{split}
	\end{equation}

	\item The coefficient vector $\boldsymbol{c}$ satisfying (\ref{rewrite form}) is computed using a general nonlinear optimization method. 
	Although the value-function approximation is linear in the coefficient vector, the HJB residual is nonlinear in this vector because the optimal control depends on the derivatives of the approximated value function. Local optimization can therefore be sensitive to the initial coefficient values.
	In this study, we first apply a genetic algorithm (GA) to search for a global candidate solution for $\boldsymbol{c}$, and then use this as the initial value for a quasi-Newton method to derive the optimal $\boldsymbol{c}$.
\end{enumerate}

\section{HJB equation residual in the benchmark}
\label{app 3}

\begin{figure}[th]
	\centering
	\includegraphics[bb=0 0 283 226, width=0.75\textwidth]{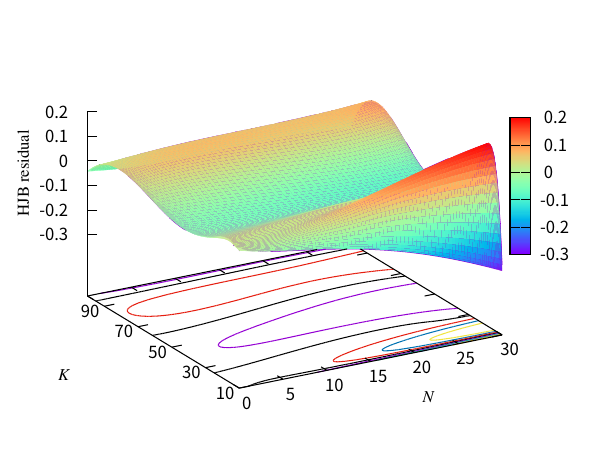}
	\caption{HJB residual in the benchmark}
	\label{fig:case0 residual}
\end{figure}

The Chebyshev polynomials employed in the numerical solution of the HJB equation are known for their ability to approximate continuous functions efficiently and with high accuracy; however, it is also well established that approximation accuracy tends to deteriorate near the boundaries of the interval over which the state variables are estimated \cite{miranda2002}. 
To verify this, Figure \ref{fig:case0 residual} presents the residuals of the HJB equation (\ref{HJB simple}) over the $(K, N)$ state space used in this analysis.
In our benchmark computation, relatively large residuals arise in the region close to the lower bound of $K$ and simultaneously close to the upper bound of $N$. Overall, however, the results demonstrate that the approximation is satisfactory.

\clearpage



\bibliographystyle{elsarticle-harv} 
\bibliography{Ishikura2026Oct}







\end{document}